\documentclass[trackchanges]{aastex62}
 \usepackage{graphicx}
 \graphicspath{ {./Images/} }
 \usepackage{subfigure}
 \usepackage{natbib}
 \usepackage{times}
 \usepackage{amsmath}
 \usepackage{textcomp}
 \usepackage{longtable}
 \usepackage{mathptmx}
 \usepackage{xcolor}
 \usepackage{booktabs}
 \usepackage[colorinlistoftodos]{todonotes}
 \usepackage{lineno}
 \usepackage{balance}

\submitjournal{ApJ}
\shorttitle{photo-z}
\shortauthors{Sheng et al.}

\def\asec{\ifmmode^{\prime\prime}\else$^{\prime\prime}$\fi}
\def\degs{\ifmmode ^{\circ}\else$^{\circ}$\fi}

\begin{document}

\title{Revealing high-$z$ \textit{Fermi}$-$LAT BL Lacs using \emph{Swift} and SARA data with photometric analysis}
\email{sheng2@clemson.edu}
\author[0000-0002-3833-1054]{Y. Sheng}
\affil{Department of Physics and Astronomy, Clemson University, SC 29634-0978, U.S.A.\footnote{A member of the SARA Consortium}}

\author[0000-0002-8979-5254]{M. Rajagopal}
\affil{Department of Physics and Astronomy, Clemson University, SC 29634-0978, U.S.A.\footnote{A member of the SARA Consortium}}

\author[0000-0002-0878-1193]{A. Kaur}
\affil{Department of Astronomy and Astrophysics, 525 Davey Lab, Pennsylvania State University, University Park, PA 16802, USA}

\author[0000-0002-6584-1703]{M. Ajello}
\affil{Department of Physics and Astronomy, Clemson University, SC 29634-0978, U.S.A.\footnote{A member of the SARA Consortium}}

\author[0000-0002-3433-4610]{A. Domínguez}
\affil{IPARCOS and Department of EMFTEL, Universidad Complutense de Madrid, E-28040 Madrid, Spain}

\author[0000-0001-5990-6243]{A. Rau}
\affil{Max-Planck-Institut für extraterrestrische Physik, Giessenbachstraße 1, D-85748 Garching, Germany}

\author{S. B. Cenko}
\affil{Astrophysics Science Division, NASA Goddard Space Flight Center, Mail Code 661, Greenbelt, MD 20771, USA}
\affil{Joint Space-Science Institute, University of Maryland, College Park, MD 20742, USA}

\author{J. Greiner}
\affil{Max-Planck-Institut für extraterrestrische Physik, Giessenbachstraße 1, D-85748 Garching, Germany}

\author[0000-0002-8028-0991]{D. H. Hartmann}
\affil{Department of Physics and Astronomy, Clemson University, SC 29634-0978, U.S.A.\footnote{A member of the SARA Consortium}}
\affil{Southeastern Association for Research in Astronomy (SARA), USA}

\author{I. Cox}
\affil{Department of Physics and Astronomy, Clemson University, SC 29634-0978, U.S.A.\footnote{A member of the SARA Consortium}}

\author[0000-0001-9427-2944]{S. Joffre}
\affil{Department of Physics and Astronomy, Clemson University, SC 29634-0978, U.S.A.\footnote{A member of the SARA Consortium}}

\author{C. Karwin}
\affil{Department of Physics and Astronomy, Clemson University, SC 29634-0978, U.S.A.\footnote{A member of the SARA Consortium}}

\author[0000-0002-8436-1254]{A. McDaniel}
\affil{Department of Physics and Astronomy, Clemson University, SC 29634-0978, U.S.A.\footnote{A member of the SARA Consortium}}

\author{R. Silver}
\affil{Department of Physics and Astronomy, Clemson University, SC 29634-0978, U.S.A.\footnote{A member of the SARA Consortium}}

\author{N. Torres-Albà}
\affil{Department of Physics and Astronomy, Clemson University, SC 29634-0978, U.S.A.\footnote{A member of the SARA Consortium}}

\begin{abstract}
\noindent BL Lacertae (BL Lac) objects are a subclass of blazar, distinguished by their featureless optical spectrum. The featureless spectrum presents a challenge in measuring the redshift of the BL Lacs. In this paper, we measure the redshift of BL Lacs using the photometric dropout technique. 
The space-based telescope \emph{Swift} and the ground-based SARA telescopes are employed to provide magnitudes in the $uvw2,\ uvm2,\ uvw1,\ u,\ b,\ v,\ g',\ r',\ i',\ z'$ filters. We observe 60 BL Lacs and report reliable redshift upper limits for 41 of them. We discover three new high-$z$ BL Lacs ($z>1.3$) at $1.74_{-0.08}^{+0.05}$, $1.88_{-0.03}^{+0.07}$, and $2.10_{-0.04}^{+0.03}$, bringing the number of high-$z$ BL Lacs found by this method up to 19. Discussions are made on the implications for the blazar sequence, the \emph{Fermi} blazar divide, and the gamma-ray horizon based on an analysis of the 4LAC catalog and all high-$z$ BL Lacs found with the photo-$z$ technique.
\end{abstract}


\section{Introduction} \label{sec:intro}

Active galactic nuclei (AGNs) are laboratories for extreme physical  processes that generate high luminosities \citep{urryUnifiedSchemesRadioLoud1995}. They are supermassive black holes at the center of galaxies \citep{fabianTestingAGNParadigm2008}, powered by active mass accretion \citep{marconiLocalSupermassiveBlack2004}. The classification of AGN is based on their orientation relative to our line of sight \citep{urryUnifiedSchemesRadioLoud1995}. Blazars are a subclass of jetted AGN oriented with the relativistic jet pointing along the line of sight with a viewing angle $\theta_m <1/\Gamma$, where $\Gamma$ is the bulk Lorentz factor \citep{Blandford_1978,Marcotulli_2017}. Blazars produce high-energy gamma-ray emission and show high variability \citep*{perlmanActiveGalacticNuclei2013}. There are two characteristic bumps in the spectral energy distribution (SED) of blazars: the bump at lower energies (infrared to X-ray) is theorized to originate from synchrotron emission produced by electrons, and the bump at higher energies (X-ray to $\gamma$-ray) is interpreted as being due to inverse Compton scattering off either the synchrotron photons \citep{maraschiBroadBandEnergy1994} or a circumnuclear photon field \citep{Dermer_1994_EC_IC}.

Blazars can be classified into different groups by their spectral properties \citep{padovaniConnectionXRayRadioselected1995}. Flat spectrum radio quasars (FSRQs) are blazars with broad emission lines in their spectra (equivalent width $>$ 5 \AA ), while BL Lacs exhibit no or weak emission lines (equivalent width $\leq$ 5 \AA) \citep{maraschiBroadBandEnergy1994,Ajello_2013}. The lack of prominent emission lines of BL Lacs makes it challenging to measure their redshifts. Indeed, around 40 \% of BL Lacs in the Fourth LAT AGN Catalog Data Release 2 (4LAC-DR2) lack measured redshifts \citep{Ajello_2020,4lac_dr2}. Additionally, blazars can be classified according to the synchrotron peak frequency: low synchrotron peaked (LSP) blazars have a synchrotron bump at lower energies where $\nu^S_{\mathrm{peak}}\leq10^{14}\ \mathrm{Hz}$; intermediate synchrotron peaked (ISP) blazars have a synchrotron bump at intermediate energies where $10^{14}\leq\nu^S_{\mathrm{peak}}\leq10^{15}\ \mathrm{Hz}$; high synchrotron peaked (HSP) blazars have a synchrotron bump at higher energies where $\nu^S_{\mathrm{peak}}\geq10^{15}\ \mathrm{Hz}$ \citep{abdoSpectralEnergyDistribution2010}.

Measuring redshifts of distant sources is essential in studying their energetic jets as well as a means of measuring the extragalactic background light (EBL, \citealt{Ackermann_2012}). The EBL is the cumulative radiation from star formation and super-massive black hole
accretion integrated over all redshifts. \citep{dominguezSpectralAnalysisFermiLAT2015,ebl_desai2019gev}. Direct measurements of the EBL are difficult because of the strong zodiacal light and foreground emission from our Galaxy \citep{moralejoMeasurementEBLCombined2017}. However, blazars can be utilized to measure the EBL as an alternative method to the direct measurement. The annihilation between $\gamma$-ray photons from blazars and EBL photons leaves a distinct attenuation in the spectra of blazars. This attenuation can be used to constrain the EBL models and its evolution with redshift \citep{Biteau_2015, Abdollahi_EBL_2018}. Moreover, the attenuation increases as the redshift increases, motivating us to search for high-redshift sources.

Photometric redshift measurement has been applied to powerlaw sources like GRBs \citep{tagliaferri2005grb, 2011Kruhler}. \cite{Rau2012} unitized the photometric method to measure the redshifts of blazars by fitting SED templates to multi-band photometry obtained by simultaneous \emph{Swift}/UVOT \citep{roming2005swift} and Gamma-Ray burst Optical/NearInfrared Detector (GROND, \citealt{greiner2007grond}) optical imaging. Photons from blazars are absorbed by the neutral hydrogen along the line of sight with a rest-wavelength blue-wards of 912\r{A} \ (Lyman-limit) and 1216\r{A}\ (Lyman-$\alpha$ forest), leaving two characteristic breaks in the source spectra. The breaks are redshift dependent and, therefore, can be used to measure the photometric redshift. The photometric technique enabled \cite{Rau2012} to measure the redshifts for nine BL Lacs out of a sample of 103 from the Second LAT AGN Catalog (2LAC) \citep{ackermannSecondCatalogActive2011}. Six of them were newly discovered BL Lacs with high redshifts ($z>1.3$). Later, using the same technique except simultaneous observations, \cite{kaur2017} found photometric redshifts for five sources from a sample of forty from the Third \emph{Fermi} Large Area Telescope Source Catalog (3FGL) \citep{collaboration_fermi_2015}, and \cite{kaur2018} discovered two more high-$z$ BL Lacs. \cite{Rajagopal_2020} continued the work and found three more sources with high redshifts, bringing the number of high-$z$ BL Lacs found by this method up to 16.

Here we continue this campaign, using the Neil Gehrels \emph{Swift} Observatory \citep{gehrels2004swift}, 1.0 m SARA-RM (Southeastern Association for Research in Astronomy at Roque de los Muchachos, La Palma, Spain), and 0.6 m SARA-CT (Cerro Tololo, Chile) telescopes \citep{keel_remote_2016} to obtain magnitudes in 10 filters from ultraviolet to near infrared (nIR) ($uvw2,\ uvm2,\ uvw1,\ u,\ b,\ v,\ g',\ r',\ i',\ z'$) of the 60 selected sources. The photometric redshifts are estimated by the method developed by \cite{Rau2012}, but the UVOT and SARA observations are not always simultaneous. A flat cosmological model $\Lambda$CDM with $H_0=73\ \mathrm{km/s/Mpc}$, $\Omega_m=0.3$, $\Omega_{\Lambda}=0.7$ is used in the content of this work. The structure of this paper is organized as follows: Section 2 introduces the instruments and observational methods used. Section 3 describes the procedures and techniques in the data analysis. Section 4 describes the SED fitting technique. Section 5 reports the redshifts for the sources we analyzed, which is followed by Section 6 that discusses our findings and applications of high-$z$ BL Lacs. Section 7 is the summary of this work.

\section{Observations} \label{sec:observations}
A sample of 60 BL Lacs with no redshift is selected from the 3FGL \citep{collaboration_fermi_2015} and Third \emph{Fermi}-LAT Catalog of High-Energy Sources (3FHL, \citealt{ajello20173fhl}). Of the 60 selected BL Lacs, 12 were observed during cycle 13\footnote{Proposal: 1316180, PI: Dr. A. Kaur.}, 16\footnote{Proposal: 1619148, PI: Dr. M. Rajagopal}, and 17\footnote{Proposal: 1720112, PI: Dr. M. Rajagopal} of the Neil Gehrels \emph{Swift} Observatory Guest Investigator program. The rest were approved as Targets of Opportunity. The selected BL Lacs are observed in the UV-optical band with six filters ($uvw2, uvm2, uvw1, u, b, v$) on the UVOT instrument of \emph{Swift}. The total exposure time of the UVOT instrument is $\sim 2000$s for each source. The time assigned to each filter is weighted as $uvw2:uvm2:uvw1:u:b:v = 4:3:2:1:1:1$

The selected targets are observed in the optical-nIR band from the ground using four SDSS filters ($g',r',i',z'$) installed on SARA-RM and SARA-CT. The exposure in every filter ranges from 15 to 40 minutes to ensure a good signal to noise ratio of the images. The time separation between the first and last filter is around 50 minutes to 1 hour and 30 minutes depending on the brightness of the sources. Table \ref{tab:obs} shows the details of the observation of our sample.

\section{Data analysis and calibrations}

\subsection{SARA data analysis}
\begin{figure}[h]
    \centering
    \includegraphics[width=0.4\textwidth]{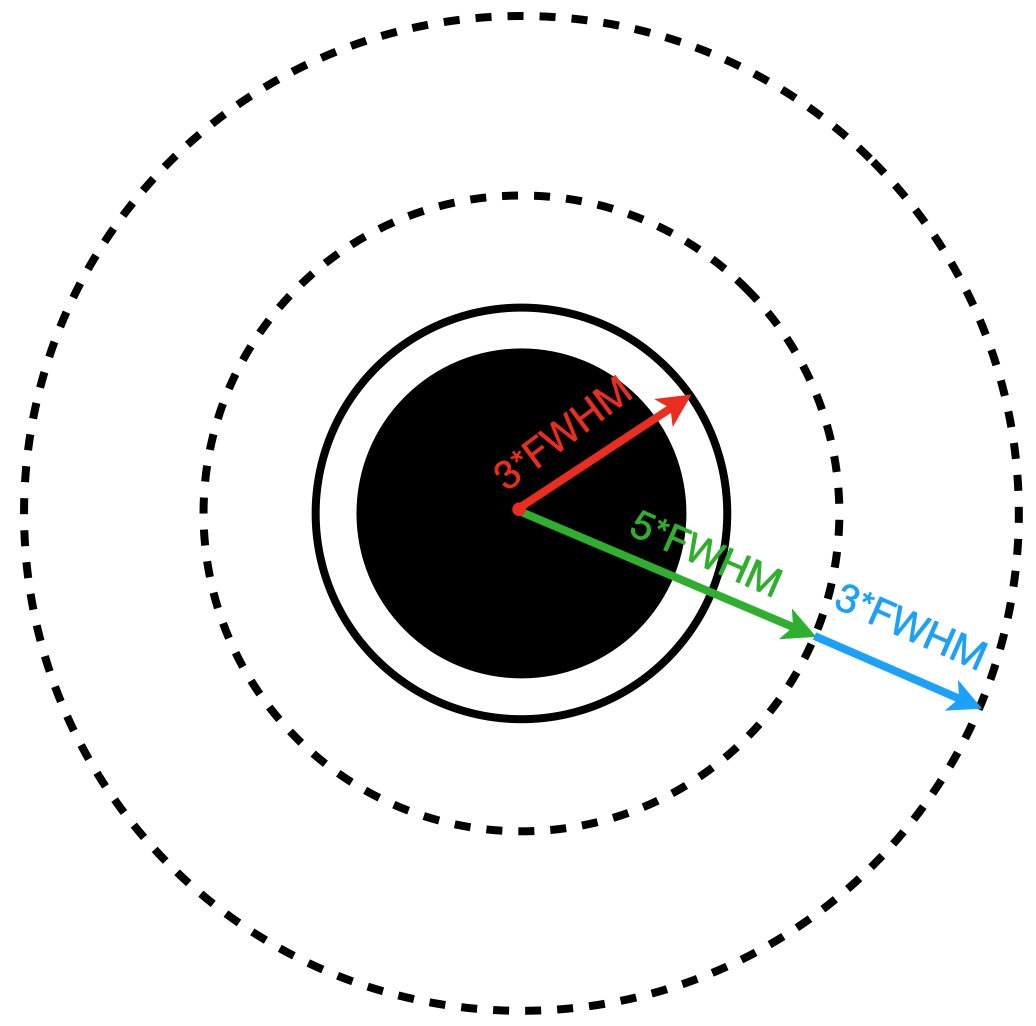}
    \caption{A circular region with radius of 3 times the FWHM is used as the source region. An annulus with inner radius of 5 times the FWHM and width of 3 times the FWHM is used as the background region. This setting makes the area of the background annulus $\sim$4 times the area of the aperture, which ensures good statistics on the background estimation.}
    \label{fig:IRAF}
\end{figure}

We calibrate the images with bias, dark and flat frames using IRAF v2.61 (Image Reduction and Analysis Facility, \citealt{todd_iraf}): the bias frames remove the electronic background, the dark frames remove the current caused by thermal effect, and the flat frames correct pixel-to-pixel variations in the sensitivity of the CCD. The same IRAF software is also used to perform photometry. We use standard stars \citep{smith2006southern,landolt_ubvri_2009,albareti_13th_2017} observed on the same night to calibrate the magnitudes of the sources. We measure the full width half maximum (FWHM) of the point-spread function for each night. A schematic representation of the source and background regions chosen for analysis are shown in Figure \ref{fig:IRAF}.

\subsection{Swift UVOT data analysis}
We analyze the UVOT data following the standard pipeline from \cite{poole_photometric_2008}. The images are calibrated using the remote \emph{Swift} Calibration Database (CALDB) v.1.0.2 to remove bad pixels, correct sensitivities, apply dark frames, etc. A circular region of 5\arcsec\ is used as the source region, and 20\arcsec \ to 30\arcsec\ is used as the background region. We use the command \texttt{UVOTSOURCE} provided in HEASoft v.6.28\footnote{https://heasarc.gsfc.nasa.gov/docs/software/heasoft/} to extract the AB magnitudes from the images. If the image file contains multiple extensions, \texttt{UVOTIMSUM} is used to align and stack the extensions. Both SARA and \emph{Swift} magnitudes are corrected for the Galactic foreground extinction \citep{kataoka_multiwavelength_2008}.

\subsection{Variability correction and cross-calibration}

Blazars are highly variable in the optical band on timescales from minutes to days \citep{racine1970photometry,carini1991timescales,urry1993multiwavelength,variability2022statistical}. Since both \emph{Swift} and SARA telescopes observe the sources in sequence, we must account for the variability caused by the non-simultaneous observation as a part of the systematic uncertainty in our photo-$z$ technique. Following \cite{Rau2012} and \cite{Rajagopal_2020}, we include a systematic uncertainty $\Delta = 0.1$ mag for both UVOT and SARA filters based on the changes in the observed magnitudes in the $uvw2$ filter.

Cross-calibration between UVOT and SARA filters is also performed following \cite{Rau2012}. We assume that the SED of the sources of interest remains unchanged. We approximate the SED by a power law. Color terms $g'-r'$ and $b-g'$ are fitted using a quadratic equation \citep{Rajagopal_2020}:
\begin{equation} \label{eq1}
b - g' = 0.26(g' - r') + 0.02(g' - r')^2.
\end{equation}
Using Equation \ref{eq1}, we calculate the offset in the $b$ filter. This offset is applied to all UVOT filters. Cross-calibrated magnitudes are reported in Table \ref{tab:mags}.

\section{SED Fitting}
The FORTRAN code LePhare v.2.2 \citep{arnouts1999measuring,ilbert2006accurate} is used to perform SED fitting and find the photometric redshifts. LePhare utilizes a $\chi^2$ fitting technique to determine a best-fit SED model for each source, based on the observational data in 10 filters. Three different libraries are included during this fitting process. The first library contains 60 power-law ($F_{\lambda} \propto \lambda^{-\beta}, \beta \in  [0, 3]$) templates. The second library consists of templates of normal galaxies and AGN/galaxy hybrids \citep{salvato2008photometric,salvato2011dissecting}, which are included to model host-galaxy-dominated sources. The third library is composed of various stellar templates to avoid false associations \citep{bohlin1995white,pickles1998stellar,chabrier2000evolutionary}.

\startlongtable
\begin{deluxetable*}{llccllc}
	\tablecolumns{7}
	\tablecaption{\label{tab:obs} $\emph{Swift}$-UVOT and SARA Observations along with visual extinction values. $A_{V}=3.1\times$E(B-V)}
	\tablewidth{0pt}
	\tabletypesize{\small}
	\setlength{\tabcolsep}{0.07in} 
	\tablehead{
		\colhead{3FGL or 3FHL}   & \colhead{$\emph{$\emph{Counterpart}$}$} & \colhead{RA J2000$^b$} &	\colhead{Dec J2000$^b$} &	\colhead{$\emph{Swift}$ Date$^{a}$} & 	\colhead{SARA Date$^{a}$} & 	\colhead{$A_{V}$} \\
		\colhead{ (Name)}   & \colhead{(Name)} & \colhead{(hh:mm:ss)} &	\colhead{($^\circ: ^\prime: ''$)} &	\colhead{(UT)} & 	\colhead{(UT)} & 	\colhead{(mag)}
    	}
\startdata
3FGL J0009.3$+$5030 & NVSS J000922$+$503028 & 00:09:22.76 & $+$50:30:28.84 & 2018 Sep. 14 & 2021 Sep. 18 & 0.40\\
3FGL J0018.9$-$8152 & PMN J0019$-$8152 & 00:19:20.64 & $-$81:52:51.42 & 2019 Nov. 13 & 2021 Jul. 27 & 0.38\\
3FGL J0021.6$-$6835 & PKS 0021$-$686 & 00:24:06.72 & $-$68:20:54.59 & 2020 Oct. 21 & 2021 Jul. 29 & 0.07\\
3FGL J0244.4$-$8224 & PMN J0251$-$8226 & 02:51:09.23 & $-$82:26:29.24 & 2020 Oct. 22 & 2021 Jul. 29 & 0.26\\
3FHL J0301.4$-$5618 & WISEA J030115.09$-$561643.5 & 03:01:14.72$^c$ & $-$56:16:44.80$^c$ & 2021 Nov. 24 & 2021 Dec. 03 & 0.03\\
3FGL J0303.6$+$4716 & 4C $+$47.08 & 03:03:35.24 & $+$47:16:16.28 & 2020 Jul. 21 & 2021 Sep. 18 & 0.68\\
3FHL J0319.2$-$7045 & WISEA J032009.21$-$704533.6 & 03:20:07.32$^c$ & $-$70:43:20.10$^c$ & 2021 Dec. 24 & 2021 Dec. 23 & 0.09\\
3FHL J0500.6$+$1903 & $...$ & 05:00:42.98$^c$ & $+$19:03:14.80$^c$ & 2021 Dec. 10 & 2022 Jan. 04 & 1.36\\
3FHL J0501.0$+$2425 & 1RXS J050107.1$+$242318 & 05:01:06.97$^c$ & $+$24:23:16.32$^c$ & 2021 Dec. 31 & 2021 Dec. 23 & 1.37\\
3FGL J0706.5$+$3744 & GB6 J0706$+$3744 & 07:06:31.70 & $+$37:44:36.42 & 2015 May 18 & 2020 Oct. 28 & 0.19\\
3FGL J0707.0$+$7741 & NVSS J070651$+$774137 & 07:06:51.33 & $+$77:41:37.00 & 2020 Apr. 12 & 2021 Feb. 16 & 0.13\\
3FGL J0753.1$+$5353 & 4C $+$54.15 & 07:53:01.38 & $+$53:52:59.63 & 2020 Aug. 18 & 2020 Oct. 28 & 0.11\\
3FGL J0754.4$-$1148 & TXS 0752$-$116 & 07:54:26.46 & $-$11:47:16.94 & 2020 Nov. 22 & 2021 Feb. 01 & 0.53\\
3FGL J0806.6$+$5933 & SBS 0802$+$596 & 08:06:25.95 & $+$59:31:06.92 & 2020 Jun. 10 & 2021 Feb. 01 & 0.16\\
3FGL J0816.7$+$5739 & SBS 0812$+$578 & 08:16:22.73 & $+$57:39:09.14 & 2022 May 31 & 2022 Apr. 09 & 0.14\\
3FGL J0839.6$+$1803 & TXS 0836$+$182 & 08:39:30.72 & $+$18:02:47.15 & 2021 Mar. 07 & 2022 Mar. 06 & 0.07\\
3FGL J0843.9$+$5311 & NVSS J084411$+$531250 & 08:44:11.70 & $+$53:12:50.58 & 2021 Mar. 30 & 2022 Mar. 11 & 0.07\\
3FGL J0849.3$+$0458 & TXS 0846$+$051 & 08:49:32.55 & $+$04:55:07.86 & 2021 Mar. 07 & 2021 Apr. 10 & 0.17\\
3FGL J0924.2$+$0534 & RBS 0771 & 09:24:01.26 & $+$05:33:42.70 & 2021 Oct. 15 & 2022 Jan.04 & 0.12\\
3FGL J0942.1$-$0756 & PMN J0942$-$0800 & 09:42:21.46 & $-$07:59:53.20 & 2018 Dec. 19 & 2022 Feb. 13 & 0.08\\
3FGL J1027.7$+$6316 & RX J1027.4$+$6317 & 10:27:25.12 & $+$63:17:53.20 & 2020 Jul. 19 & 2021 Feb. 01 & 0.03\\
3FGL J1032.5$+$6623 & SDSS J103239.06$+$662323.2 & 10:32:39.08 & $+$66:23:23.21 & 2020 Apr. 03 & 2021 Feb. 16 & 0.04\\
3FGL J1101.5$+$4106 & RX J1101.3$+$4108 & 11:01:24.73 & $+$41:08:47.39 & 2021 Oct. 04 & 2022 Mar. 11 & 0.03\\
3FGL J1103.9$-$5357 & PKS 1101$-$536 & 11:03:52.22 & $-$53:57:00.68 & 2021 Oct. 16 & 2021 Dec. 23 & 0.76\\
3FHL J1127.8$+$3615 & WISE J112758.88$+$362028.4 & 11:27:59.08$^c$ & $+$36:20:33.09$^c$ & 2021 Jul. 18 & 2022 Mar. 11 & 0.06\\
3FGL J1224.6$+$4332 & B3 1222$+$438 & 12:24:51.51 & $+$43:35:19.28 & 2020 Nov. 15 & 2022 May 26 & 0.03\\
3FGL J1244.8$+$5707 & 1RXS J124510.5$+$571020 & 12:45:09.99 & $+$57:09:54.36 & 2020 Nov. 30 & 2021 Feb. 16 & 0.03\\
3FGL J1249.7$+$3705 & RX J1249.8$+$3708 & 12:49:46.75 & $+$37:07:47.90 & 2021 May 06 & 2021 Jun. 12 & 0.04\\
3FGL J1256.7$+$5328 & TXS 1254$+$538 & 12:56:38.630 & $+$53:34:23.70 & 2020 Dec. 15 & 2021 Jun. 27 & 0.05\\
3FGL J1311.0$+$0036 & RX J1311.1$+$0035 & 13:11:06.48 & $+$00:35:10.03 & 2022 Apr. 13 & 2022 Feb. 26 & 0.08\\
3FGL J1327.9$+$2524 & NVSS J132758$+$252750 & 13:27:58.93 & $+$25:27:46.71 & 2021 Apr. 13 & 2022 Apr. 09 & 0.03\\
3FGL J1357.5$+$0125 & RX J1357.6$+$0128 & 13:57:38.70 & $+$01:28:13.62 & 2022 Apr. 26 & 2022 Apr. 21 & 0.09\\
3FGL J1419.5$+$0449 & SDSS J141927.49$+$044513.7 & 14:19:27.74 & $+$04:45:13.93 & 2022 Apr. 23 & 2022 Apr. 21 & 0.09\\
3FHL J1421.5$-$1654 & WISEA J142128.94$-$165455.4 & 14:21:29.22$^c$ & $-$16:54:55.13$^c$ & 2021 Aug. 29 & 2022 Feb. 13 & 0.24\\
3FGL J1424.3$+$0434 & TXS 1421$+$048 & 14:24:09.50 & $+$04:34:52.07 & 2022 Mar. 28 & 2022 Feb. 26 & 0.07\\
3FGL J1426.2$+$3402 & RGB J1426$+$340 & 14:26:07.72 & $+$34:04:26.33 & 2021 Apr. 10 & 2022 May 26 & 0.05\\
3FHL J1440.2$-$2343 & WISEA J143959.46$-$234141.0 & 14:39:59.46$^c$ & $-$23:41:41.0$^c$ & 2021 Sep. 15 & 2021 Aug. 20 & 0.30\\
3FGL J1446.1$-$1628 & PKS B1443$-$162 & 14:45:53.38 & $-$16:29:01.61 & 2022 Jan. 03 & 2022 Feb. 26 & 0.29\\
3FGL J1454.5$+$5124 & TXS 1452$+$516 & 14:54:27.12 & $+$51:24:33.73 & 2021 Jul. 01 & 2022 May 06 & 0.04\\
3FGL J1509.7$+$5556 & SBS 1508$+$561 & 15:09:47.95 & $+$55:56:17.30 & 2021 May 17 & 2021 Jul. 02 & 0.04\\
3FGL J1531.0$+$5737 & 87GB 152947.5$+$574636 & 15:30:58.34 & $+$57:36:25.06 & 2021 May 22 & 2022 May 06 & 0.04\\
3FGL J1533.5$+$3416 & RX J1533.3$+$3416 & 15:33:24.26 & $+$34:16:40.22 & 2021 May 24 & 2021 Aug. 07 & 0.06\\
3FGL J1534.4$+$5323 & 1ES 1533$+$535 & 15:35:00.81 & $+$53:20:37.21 & 2021 May 31 & 2022 May 06 & 0.05\\
3FGL J1550.3$+$7409 & 87GB 155014.9$+$741816 & 15:49:27.28 & $+$74:09:32.18 & 2020 Sep. 29 & 2021 Jun. 12 & 0.07\\
3FGL J1549.0$+$6309 & SDSS J154958.45$+$631021.2 & 15:49:57.32 & $+$63:10:7.32 & 2021 Jan. 07 & 2021 Jun. 12 & 0.05\\
3FGL J1630.7$+$5222 & TXS 1629$+$524 & 16:30:43.15 & $+$52:21:38.63 & 2021 Jul. 10 & 2021 Jul. 02 & 0.07\\
3FGL J1637.1$+$1314 & 1RXS J163717.1$+$131418 & 16:37:16.74 & $+$13:14:38.80 & 2021 Jul. 17 & 2022 May 06 & 0.16\\
3FGL J1651.6$+$7219 & RX J1651.6$+$7218 & 16:51:39.95 & $+$72:18:25.13 & 2022 Apr. 17 & 2022 May 26 & 0.11\\
3FGL J1829.8$+$1328 & MG1 J183001$+$1323 & 18:30:00.76 & $+$13:24:14.33 & 2021 Nov. 11 & 2022 May 06 & 0.67\\
3FGL J1848.9$+$4247 & RGB J1848$+$427 & 18:48:47.17 & $+$42:45:34.92 & 2021 Nov. 17 & 2022 May 06 & 0.22\\
3FGL J1927.7$+$6118 & S4 1926$+$61 & 19:27:30.44 & $+$61:17:32.87 & 2021 Sep. 06 & 2021 Sep. 18 & 0.17\\
3FGL J2007.7$-$7728 & PKS 2000$-$776 & 20:12:30.18 & $-$16:46:50.48 & 2020 Jul. 20 & 2021 Jun. 11 & 0.32\\
3FGL J2012.1$-$1643 & PMN J2012$-$1646 & 20:12:30.18 & $-$16:46:50.48 & 2020 Jul. 18 & 2020 Oct. 28 & 0.32\\
3FGL J2107.7$-$4822 & PMN J2107$-$4827 & 21:07:44.48 & $-$48:28:03.00 & 2017 Sep. 08 & 2021 Jul. 27 & 0.10\\
3FGL J2139.4$-$4235 & MH 2136$-$428 & 21:39:24.17 & $-$42:35:20.00 & 2020 Sep. 25 & 2021 Jul. 27 & 0.05\\
3FGL J2149.6$+$1915 & TXS 2147$+$191 & 21:49:47.31 & $+$19:20:46.07 & 2020 Aug. 25 & 2021 Jul. 02 & 0.31\\
3FGL J2159.8$+$1025 & TXS 2157$+$102 & 22:00:07.93 & $+$10:30:07.88 & 2020 Oct. 21 & 2020 Oct. 28 & 0.17\\
3FGL J2236.0$-$3629 & NVSS J223554$-$362901 & 22:35:54.83 & $-$36:29:02.87 & 2020 Oct. 30 & 2021 Jul. 29 & 0.04\\
3FGL J2236.2$-$5049 & SUMSS J223605$-$505521 & 22:36:05.64 & $-$50:55:19.92 & 2020 Aug. 04 & 2021 Jul. 29 & 0.04\\
3FGL J2240.9$+$4121 & B3 2238$+$410 & 22:41:07.20 & $+$41:20:11.62 & 2021 Aug. 01 & 2021 Aug. 07 & 0.65\\
\enddata
    \end{deluxetable*}

\begin{footnotesize}
$^a$ The beginning date of the exposures for Swift and SARA.

$^b$ The coordinates are from the 3FGL catalog, which are submitted to \emph{Swift} for the UVOT observation and SARA telescopes for the optical observation.

$^c$ The coordinates are determined by the method discussed in \cite{Joffre_2022, stroh2013swift,paiano2017optical,silver2020identifying,Kerby_2021}
\end{footnotesize}  

\begin{center}
\begin{longrotatetable}
\begin{deluxetable*}{lllllllllll}
	\tablecolumns{11}
	\tablecaption{\label{tab:mags} $\emph{Swift}$-UVOT and SARA photometry (AB magnitudes corrected for extinction)}
	\tablewidth{0pt}
	\tabletypesize{\footnotesize}
	\setlength{\tabcolsep}{0.03in} 
	\tablehead{
\colhead{3FGL or 3FHL Name}   & \colhead{$g^\prime$} & \colhead{$r^\prime$} &	\colhead{$i^\prime$} &	\colhead{$z^\prime$} & 	\colhead{$uvw2$} & 	\colhead{$uvm2$} & 	\colhead{$uvw1$} & 	\colhead{$u$} & \colhead{$b$} & \colhead{$v$}}
	\startdata
3FGL J0009.3$+$5030 & 18.38$\pm$0.01 & 18.06$\pm$0.01 & 17.78$\pm$0.01 & 17.50$\pm$0.02 & 20.25$\pm$0.16 & 20.15$\pm$0.19 & 19.55$\pm$0.15 & 19.10$\pm$0.13 & 18.46$\pm$0.12 & 17.78$\pm$0.13\\
3FGL J0018.9$-$8152 & 17.60$\pm$0.02 & 17.16$\pm$0.01 & 16.84$\pm$0.01 & 16.67$\pm$0.03 & 18.87$\pm$0.07 & 18.91$\pm$0.09 & 18.51$\pm$0.08 & 18.08$\pm$0.06 & 17.72$\pm$0.06 & 17.54$\pm$0.09\\
3FGL J0021.6$-$6835 & 18.02$\pm$0.01 & 17.58$\pm$0.01 & 17.52$\pm$0.01 & 16.84$\pm$0.03 & 20.03$\pm$0.07 & 19.72$\pm$0.08 & 19.35$\pm$0.07 & 18.60$\pm$0.05 & 18.14$\pm$0.05 & 18.02$\pm$0.08\\
3FGL J0244.4$-$8224 & 18.70$\pm$0.02 & 18.49$\pm$0.02 & 18.30$\pm$0.03 & 17.89$\pm$0.06 & 19.54$\pm$0.09 & 19.60$\pm$0.13 & 19.87$\pm$0.16 & 19.16$\pm$0.13 & 18.75$\pm$0.15 & 18.42$\pm$0.30\\
3FHL J0301.4$-$5618 & 19.08$\pm$0.02 & 18.56$\pm$0.01 & 18.27$\pm$0.02 & 17.53$\pm$0.04 & $>$21.91 & $>$21.42 & 20.44$\pm$0.23 & 19.46$\pm$0.18 & 19.22$\pm$0.26 & $>$18.78\\
3FGL J0303.6$+$4716 & 16.59$\pm$0.00 & 16.13$\pm$0.00 & 15.86$\pm$0.00 & 15.57$\pm$0.00 & 18.61$\pm$0.22 & 18.51$\pm$0.24 & 17.91$\pm$0.17 & 17.13$\pm$0.11 & 16.71$\pm$0.12 & 16.06$\pm$0.14\\
3FHL J0319.2$-$7045 & 19.72$\pm$0.07 & 18.96$\pm$0.03 & 18.59$\pm$0.04 & 17.60$\pm$0.05 & 21.00$\pm$0.17 & 20.74$\pm$0.22 & 20.71$\pm$0.29 & 20.52$\pm$0.36 & $>$19.92 & 18.95$\pm$0.30\\
3FHL J0500.6$+$1903 & 19.73$\pm$0.09 & 19.18$\pm$0.05 & 19.35$\pm$0.07 & 19.45$\pm$0.21 & $>$18.05 & $>$20.44 & $>$20.27 & $>$20.28 & $>$19.88 & $>$19.46\\
3FHL J0501.0$+$2425 & 18.50$\pm$0.08 & 18.05$\pm$0.04 & 17.89$\pm$0.05 & 18.26$\pm$0.16 & $>$19.11 & $>$19.24 & $>$19.01 & 18.86$\pm$0.32 & $>$18.62 & $>$18.19\\
3FGL J0706.5$+$3744 & 18.01$\pm$0.01 & 17.59$\pm$0.01 & 17.24$\pm$0.01 & 17.30$\pm$0.02 & 19.07$\pm$0.07 & 19.05$\pm$0.09 & 18.82$\pm$0.07 & 18.42$\pm$0.09 & 18.13$\pm$0.11 & 18.06$\pm$0.21\\
3FGL J0707.0$+$7741 & 17.63$\pm$0.00 & 17.08$\pm$0.00 & 16.83$\pm$0.00 & 16.50$\pm$0.01 & 18.88$\pm$0.08 & 19.06$\pm$0.11 & 18.68$\pm$0.10 & 18.01$\pm$0.08 & 17.78$\pm$0.10 & 17.62$\pm$0.16\\
3FGL J0753.1$+$5353 & 19.87$\pm$0.04 & 19.42$\pm$0.03 & 19.31$\pm$0.05 & 18.83$\pm$0.09 & 22.06$\pm$0.30 & $>$22.07 & 21.42$\pm$0.32 & $>$20.73 & $>$19.99 & $>$19.12\\
3FGL J0754.4$-$1148 & 19.98$\pm$0.04 & 19.05$\pm$0.02 & 18.75$\pm$0.03 & 18.52$\pm$0.08 & 22.69$\pm$0.35 & $>$22.47 & 22.09$\pm$0.32 & 21.33$\pm$0.24 & 20.25$\pm$0.16 & 19.93$\pm$0.22\\
3FGL J0806.6$+$5933 & 18.93$\pm$0.01 & 18.43$\pm$0.01 & 17.77$\pm$0.01 & 17.58$\pm$0.03 & 20.33$\pm$0.09 & 20.09$\pm$0.10 & 20.05$\pm$0.11 & 19.48$\pm$0.11 & 19.06$\pm$0.13 & 18.23$\pm$0.13\\
3FGL J0816.7$+$5739 & 17.03$\pm$0.01 & 16.62$\pm$0.00 & 16.34$\pm$0.01 & 16.18$\pm$0.01 & 18.24$\pm$0.07 & 18.13$\pm$0.09 & 17.84$\pm$0.09 & 17.19$\pm$0.08 & 17.13$\pm$0.11 & 16.69$\pm$0.15\\
3FGL J0839.6$+$1803 & 17.81$\pm$0.01 & 17.33$\pm$0.01 & 17.02$\pm$0.01 & 16.56$\pm$0.02 & 17.84$\pm$0.14 & 18.74$\pm$0.09 & 18.49$\pm$0.09 & 18.19$\pm$0.10 & 17.94$\pm$0.14 & 17.28$\pm$0.16\\
3FGL J0843.9$+$5311 & 17.71$\pm$0.00 & 17.42$\pm$0.00 & 17.17$\pm$0.01 & 17.05$\pm$0.01 & 19.25$\pm$0.10 & 19.23$\pm$0.13 & 18.86$\pm$0.12 & 18.36$\pm$0.12 & 17.79$\pm$0.13 & 17.53$\pm$0.20\\
3FGL J0849.3$+$0458 & 18.90$\pm$0.03 & 18.61$\pm$0.03 & 18.18$\pm$0.03 & 18.07$\pm$0.10 & $>$18.75 & 19.58$\pm$0.19 & 19.05$\pm$0.15 & 18.97$\pm$0.21 & $>$18.98 & 18.06$\pm$0.34\\
3FGL J0924.2$+$0534 & 18.20$\pm$0.01 & 17.85$\pm$0.01 & 17.65$\pm$0.01 & 17.48$\pm$0.03 & 18.85$\pm$0.07 & 18.62$\pm$0.09 & 18.81$\pm$0.09 & 18.46$\pm$0.11 & 18.29$\pm$0.17 & 17.93$\pm$0.25\\
3FGL J0942.1$-$0756 & 19.30$\pm$0.08 & 19.04$\pm$0.06 & 18.83$\pm$0.07 & 17.90$\pm$0.06 & 21.02$\pm$0.13 & 20.79$\pm$0.17 & 20.82$\pm$0.17 & 20.07$\pm$0.15 & 19.37$\pm$0.14 & 19.09$\pm$0.22\\
3FGL J1027.7$+$6316 & 19.31$\pm$0.03 & 19.17$\pm$0.03 & 18.99$\pm$0.04 & 18.81$\pm$0.09 & 20.42$\pm$0.10 & 20.24$\pm$0.12 & 20.08$\pm$0.16 & 19.52$\pm$0.17 & 19.34$\pm$0.24 & $>$19.27\\
3FGL J1032.5$+$6623 & 19.73$\pm$0.01 & 19.27$\pm$0.01 & 19.09$\pm$0.01 & 19.03$\pm$0.06 & $>$22.11 & $>$21.73 & 20.34$\pm$0.19 & 20.01$\pm$0.22 & 19.86$\pm$0.36 & 18.79$\pm$0.31\\
3FGL J1101.5$+$4106 & 18.43$\pm$0.01 & 18.28$\pm$0.01 & 18.11$\pm$0.01 & 18.07$\pm$0.03 & 19.23$\pm$0.09 & 19.00$\pm$0.11 & 18.91$\pm$0.11 & 18.77$\pm$0.18 & 18.47$\pm$0.26 & $>$18.01\\
3FGL J1103.9$-$5357 & 15.68$\pm$0.01 & 15.08$\pm$0.01 & 14.63$\pm$0.01 & 14.50$\pm$0.01 & 16.83$\pm$0.10 & 17.55$\pm$0.19 & 16.11$\pm$0.09 & 15.89$\pm$0.07 & 15.84$\pm$0.09 & 15.38$\pm$0.10\\
3FHL J1127.8$+$3615 & 20.48$\pm$0.03 & 20.34$\pm$0.03 & 20.35$\pm$0.04 & 20.22$\pm$0.11 & 21.73$\pm$0.13 & 21.48$\pm$0.14 & 21.30$\pm$0.17 & 20.79$\pm$0.24 & $>$20.51 & $>$19.70\\
3FGL J1224.6$+$4332 & 25.31$\pm$0.02 & 25.18$\pm$0.02 & 25.86$\pm$0.03 & 27.45$\pm$0.07 & $>$27.61 & $>$27.12 & $>$26.82 & $>$26.15 & $>$25.34 & $>$24.44\\
3FGL J1244.8$+$5707 & 18.91$\pm$0.01 & 18.55$\pm$0.01 & 18.31$\pm$0.01 & 17.90$\pm$0.03 & 20.11$\pm$0.07 & 19.95$\pm$0.08 & 19.75$\pm$0.08 & 19.33$\pm$0.07 & 19.01$\pm$0.08 & 18.82$\pm$0.14\\
3FGL J1249.7$+$3705 & 19.25$\pm$0.01 & 18.98$\pm$0.01 & 18.88$\pm$0.01 & 18.75$\pm$0.05 & 20.11$\pm$0.11 & 19.87$\pm$0.17 & 19.74$\pm$0.12 & 19.61$\pm$0.14 & 19.32$\pm$0.19 & $>$18.74\\
3FGL J1256.7$+$5328 & 21.84$\pm$0.08 & 20.90$\pm$0.05 & 20.46$\pm$0.09 & 19.93$\pm$0.28 & $>$24.44 & $>$24.09 & $>$23.72 & $>$22.81 & $>$22.10 & $>$21.09\\
3FGL J1311.0$+$0036 & 18.81$\pm$0.03 & 18.20$\pm$0.02 & 17.71$\pm$0.02 & 17.45$\pm$0.05 & 20.16$\pm$0.09 & 20.02$\pm$0.11 & 20.01$\pm$0.12 & 19.35$\pm$0.11 & 18.98$\pm$0.14 & $>$19.30\\
3FGL J1327.9$+$2524 & 21.40$\pm$0.08 & 20.85$\pm$0.07 & 21.26$\pm$0.22 & $...$ & $>$23.86 & $>$23.41 & $>$22.99 & $>$22.32 & $>$21.54 & $>$20.67\\
3FGL J1357.5$+$0125 & 18.08$\pm$0.01 & 17.99$\pm$0.01 & 17.55$\pm$0.01 & 16.87$\pm$0.03 & 18.90$\pm$0.08 & 18.64$\pm$0.10 & 18.64$\pm$0.09 & 18.09$\pm$0.09 & 18.11$\pm$0.14 & 17.52$\pm$0.17\\
3FGL J1419.5$+$0449 & 18.54$\pm$0.02 & 18.52$\pm$0.01 & 18.44$\pm$0.02 & 17.98$\pm$0.05 & $>$21.68 & $>$21.19 & 19.70$\pm$0.16 & 18.69$\pm$0.13 & 18.55$\pm$0.19 & $>$18.54\\
3FHL J1421.5$-$1654 & 19.30$\pm$0.04 & 18.94$\pm$0.03 & 18.75$\pm$0.04 & 17.95$\pm$0.06 & 21.04$\pm$0.26 & 20.20$\pm$0.20 & 20.67$\pm$0.21 & 20.03$\pm$0.32 & $>$19.39 & $>$18.23\\
3FGL J1424.3$+$0434 & 18.38$\pm$0.02 & 17.96$\pm$0.02 & 17.59$\pm$0.02 & 17.19$\pm$0.04 & 19.97$\pm$0.09 & 19.89$\pm$0.14 & 19.54$\pm$0.11 & 18.95$\pm$0.11 & 18.50$\pm$0.13 & 18.34$\pm$0.22\\
3FGL J1426.2$+$3402 & 22.26$\pm$0.01 & 22.44$\pm$0.01 & 23.35$\pm$0.01 & 25.47$\pm$0.02 & 23.44$\pm$0.08 & 23.23$\pm$0.09 & 23.24$\pm$0.10 & 22.53$\pm$0.09 & 22.22$\pm$0.10 & 22.28$\pm$0.20\\
3FHL J1440.2$-$2343 & 19.01$\pm$0.04 & 18.40$\pm$0.02 & 18.04$\pm$0.02 & 17.78$\pm$0.04 & 20.26$\pm$0.16 & 20.15$\pm$0.19 & 20.19$\pm$0.26 & $>$19.93 & $>$19.18 & $>$18.41\\
3FGL J1446.1$-$1628 & 20.60$\pm$0.14 & 20.04$\pm$0.07 & 19.72$\pm$0.09 & $...$ & $>$22.76 & $>$22.36 & $>$22.04 & $>$21.50 & $>$20.75 & $>$19.99\\
3FGL J1454.5$+$5124 & 16.13$\pm$0.00 & 15.74$\pm$0.00 & 15.45$\pm$0.00 & 15.26$\pm$0.00 & 17.17$\pm$0.05 & 17.05$\pm$0.06 & 16.77$\pm$0.06 & 16.46$\pm$0.05 & 16.24$\pm$0.05 & 16.02$\pm$0.08\\
3FGL J1509.7$+$5556 & 17.87$\pm$0.00 & 17.54$\pm$0.00 & 17.30$\pm$0.01 & 17.14$\pm$0.03 & 19.00$\pm$0.07 & 18.92$\pm$0.09 & 18.80$\pm$0.09 & 18.38$\pm$0.08 & 17.96$\pm$0.08 & 17.77$\pm$0.13\\
3FGL J1531.0$+$5737 & 19.68$\pm$0.01 & 19.25$\pm$0.01 & 18.96$\pm$0.02 & 18.79$\pm$0.06 & 21.77$\pm$0.16 & 21.15$\pm$0.16 & 21.12$\pm$0.17 & 20.24$\pm$0.14 & 19.80$\pm$0.16 & 21.85$\pm$0.16\\
3FGL J1533.5$+$3416 & 18.10$\pm$0.01 & 17.86$\pm$0.01 & 17.70$\pm$0.01 & 17.77$\pm$0.03 & 19.08$\pm$0.08 & 19.06$\pm$0.09 & 18.83$\pm$0.09 & 18.29$\pm$0.08 & 18.17$\pm$0.11 & 17.80$\pm$0.16\\
3FGL J1534.4$+$5323 & 18.26$\pm$0.01 & 18.00$\pm$0.01 & 17.79$\pm$0.01 & 17.68$\pm$0.02 & 18.94$\pm$0.06 & 18.73$\pm$0.07 & 18.66$\pm$0.07 & 18.47$\pm$0.07 & 18.33$\pm$0.09 & 18.21$\pm$0.15\\
3FGL J1550.3$+$7409 & 19.88$\pm$0.01 & 19.29$\pm$0.01 & 19.06$\pm$0.01 & 18.73$\pm$0.04 & 21.58$\pm$0.18 & 22.28$\pm$0.33 & 21.29$\pm$0.25 & 20.64$\pm$0.32 & $>$20.05 & $>$19.23\\
3FGL J1549.0$+$6309 & 20.40$\pm$0.02 & 19.74$\pm$0.01 & 19.22$\pm$0.02 & 19.24$\pm$0.08 & 22.11$\pm$0.25 & $>$22.06 & 21.61$\pm$0.28 & $>$21.33 & $>$20.59 & $>$19.70\\
3FGL J1630.7$+$5222 & 16.77$\pm$0.00 & 16.51$\pm$0.00 & 16.27$\pm$0.00 & 16.08$\pm$0.01 & 17.96$\pm$0.07 & 17.85$\pm$0.08 & 17.48$\pm$0.07 & 17.26$\pm$0.07 & 16.84$\pm$0.08 & 16.51$\pm$0.10\\
3FGL J1637.1$+$1314 & 19.75$\pm$0.01 & 19.42$\pm$0.01 & 19.09$\pm$0.02 & 18.65$\pm$0.06 & 21.00$\pm$0.12 & 20.69$\pm$0.14 & 20.64$\pm$0.16 & 20.05$\pm$0.16 & 19.84$\pm$0.22 & 21.26$\pm$0.12\\
3FGL J1651.6$+$7219 & 23.45$\pm$0.01 & 23.53$\pm$0.01 & 24.31$\pm$0.01 & 26.38$\pm$0.04 & 24.46$\pm$0.10 & 24.25$\pm$0.12 & 24.38$\pm$0.15 & 23.81$\pm$0.14 & 23.43$\pm$0.17 & 23.39$\pm$0.31\\
3FGL J1829.8$+$1328 & 18.06$\pm$0.01 & 17.52$\pm$0.01 & 17.11$\pm$0.01 & 17.15$\pm$0.01 & $>$20.23 & $>$19.93 & 19.30$\pm$0.26 & 19.28$\pm$0.35 & 18.20$\pm$0.23 & $>$21.35\\
3FGL J1848.9$+$4247 & 19.16$\pm$0.01 & 18.87$\pm$0.01 & 18.61$\pm$0.02 & 18.35$\pm$0.05 & 19.88$\pm$0.11 & 19.80$\pm$0.15 & 19.74$\pm$0.18 & 19.40$\pm$0.26 & $>$19.24 & 20.23$\pm$0.11\\
3FGL J1927.7$+$6118 & 18.09$\pm$0.01 & 17.58$\pm$0.01 & 17.25$\pm$0.01 & 16.92$\pm$0.01 & 19.62$\pm$0.09 & 19.43$\pm$0.10 & 19.18$\pm$0.09 & 18.61$\pm$0.08 & 18.23$\pm$0.08 & 18.02$\pm$0.13\\
3FGL J2007.7$-$7728 & 20.26$\pm$0.06 & 19.70$\pm$0.04 & 19.61$\pm$0.08 & 19.26$\pm$0.19 & $>$22.01 & $>$21.73 & $>$21.48 & $>$21.08 & $>$20.41 & $>$19.64\\
3FGL J2012.1$-$1643 & 20.01$\pm$0.06 & 19.98$\pm$0.07 & 19.72$\pm$0.06 & 20.13$\pm$0.26 & $>$22.27 & $>$22.03 & $>$21.48 & $>$20.79 & $>$20.02 & $>$19.26\\
3FGL J2107.7$-$4822 & 20.09$\pm$0.07 & 19.29$\pm$0.03 & 18.71$\pm$0.03 & 18.38$\pm$0.08 & 21.87$\pm$0.22 & 21.78$\pm$0.27 & $>$21.73 & $>$21.06 & $>$20.30 & $>$19.45\\
3FGL J2139.4$-$4235 & 16.40$\pm$0.01 & 15.86$\pm$0.00 & 15.57$\pm$0.00 & 15.31$\pm$0.01 & 17.76$\pm$0.04 & 17.61$\pm$0.05 & 17.38$\pm$0.05 & 16.98$\pm$0.04 & 16.55$\pm$0.04 & 16.22$\pm$0.05\\
3FGL J2149.6$+$1915 & 20.63$\pm$0.04 & 20.33$\pm$0.04 & 20.39$\pm$0.09 & 21.06$\pm$0.57 & 21.33$\pm$0.15 & 21.88$\pm$0.29 & 21.48$\pm$0.26 & 20.70$\pm$0.20 & $>$20.71 & $>$19.95\\
3FGL J2159.8$+$1025 & 19.17$\pm$0.03 & 18.68$\pm$0.02 & 18.29$\pm$0.02 & 18.13$\pm$0.04 & $>$21.44 & 20.36$\pm$0.30 & 20.63$\pm$0.35 & $>$20.06 & $>$19.30 & $>$18.49\\
3FGL J2236.0$-$3629 & 19.00$\pm$0.02 & 18.42$\pm$0.01 & 18.32$\pm$0.02 & 18.20$\pm$0.06 & 19.11$\pm$0.07 & 21.38$\pm$0.15 & 21.16$\pm$0.12 & 20.10$\pm$0.09 & 19.16$\pm$0.07 & 18.97$\pm$0.12\\
3FGL J2236.2$-$5049 & 18.65$\pm$0.02 & 18.39$\pm$0.01 & 18.32$\pm$0.02 & 17.98$\pm$0.05 & $>$20.98 & $>$20.40 & 20.16$\pm$0.35 & $>$19.49 & $>$18.72 & $>$17.83\\
3FGL J2240.9$+$4121 & 17.92$\pm$0.01 & 17.54$\pm$0.01 & 17.29$\pm$0.01 & 17.14$\pm$0.02 & 19.43$\pm$0.22 & $>$19.73 & 18.76$\pm$0.18 & 18.44$\pm$0.16 & 18.03$\pm$0.17 & 17.87$\pm$0.29\\
\enddata
\end{deluxetable*}
\end{longrotatetable}
\end{center}


\startlongtable
\begin{deluxetable*}{ll|ccll|lcll}
\centering
	\tablecolumns{10}
	\tabletypesize{\small}
	\tablewidth{0pt}
	\tablecaption{\label{tab:result}SED fitting}
	\setlength{\tabcolsep}{0.05in} 
	\tablehead{
		\colhead{3FGL or 3FHL Name}  & \colhead{$z_{\rm phot, best}^a$}   & \multicolumn{4}{c}{Power Law Template} & \multicolumn{4}{c}{Galaxy Template} \\
		\cmidrule[\heavyrulewidth](lr){3-6}   \cmidrule[\heavyrulewidth](lr){7-10}
		\colhead{} & \colhead{} & \colhead{$z_{\rm phot}  ^b$} & \colhead{$\chi^{2\ c}$} & \colhead{$P_{\rm z}^d$} & \colhead{$\beta^e$} \hspace{0.4cm} & \colhead{$z_{\rm phot}  ^b$} & \colhead{$\chi^{2\ c}$} & \colhead{$P_{\rm z}^d$} & \colhead{Model}} 
    \startdata
	\multicolumn{10}{c}{\textbf{\small Sources with confirmed photometric redshifts}} \\
	\hline
3FHL J0301.4$-$5618 & $1.74^{+0.05}_{-0.08}$ & $1.74^{+0.05}_{-0.08}$ & 8.1 & 95.5 & 1.90 & $1.33^{+0.06}_{-0.05}$ & 7.9 & 99.4 & \\ 
3FGL J1032.5$+$6623 & $1.88^{+0.07}_{-0.03}$ & $1.88^{+0.07}_{-0.03}$ & 8.7 & 98.5 & 1.00 & $1.49^{+0.39}_{-0.01}$ & 13.6 & 36.5 & I22491\_40\_TQSO1\_60.sed\\
3FGL J1419.5$+$0449 & $2.10^{+0.03}_{-0.04}$ & $2.10^{+0.03}_{-0.04}$ & 22.4 & 100.0 & 0.95 & $1.73^{+0.01}_{-0.04}$ & 3.6 & 96.8 & CB1\_0\_LOIII4.sed\\
	\hline
	\multicolumn{10}{c}{\textbf{\small Sources with photometric redshift upper limits}}\\
	\hline
3FGL J0009.3$+$5030 & $<1.65$ & $1.53^{+0.12}_{-0.19}$ & 10.2 & 63.9 & 1.40 & $0.00^{+0.02}_{-0.00}$ & 10.5 & 88.3 & Spi4\_template\_norm.sed\\
3FGL J0018.9$-$8152 & $<1.03$ & $0.79^{+0.24}_{-0.79}$ & 4.2 & 32.2 & 1.40 & $0.00^{+0.01}_{-0.00}$ & 6.4 & 98.2 & I22491\_80\_TQSO1\_20.sed\\
3FGL J0021.6$-$6835 & $<1.50$ & $1.42^{+0.08}_{-0.18}$ & 13.6 & 72.8 & 1.55 & $0.00^{+0.01}_{-0.00}$ & 34.3 & 99.1 & Spi4\_template\_norm.sed\\
3FGL J0244.4$-$8224 & $<0.92$ & $0.06^{+0.86}_{-0.06}$ & 11.3 & 17.7 & 1.10 & $0.04^{+0.02}_{-0.03}$ & 8.9 & 69.6 & I22491\_60\_TQSO1\_40.sed\\
3FGL J0303.6$+$4716 & $<1.47$ & $1.36^{+0.11}_{-0.08}$ & 7.3 & 65.5 & 1.60 & $0.00^{+0.04}_{-0.00}$ & 10.5 & 96.5 & Mrk231\_template\_norm.sed\\
3FHL J0319.2$-$7045 & $...$ & $3.06^{+0.02}_{-0.00}$ & 82.0 & 99.3 & 1.70 & $0.64^{+0.13}_{-0.05}$ & 22.7 & 65.7 & Mrk231\_template\_norm.sed\\
3FHL J0500.6$+$1903$^f$ & $...$ & $...$ & $...$ & $...$ & $...$ & $2.97^{+0.04}_{-2.97}$ & 1.6 & 78.9 & pl\_QSO\_DR2\_029\_t0.spec\\
3FHL J0501.0$+$2425 & $...$ & $2.82^{+0.13}_{-0.31}$ & 9.3 & 41.3 & 0.60 & $2.84^{+0.15}_{-2.84}$ & 8.8 & 87.0 & pl\_QSO\_DR2\_029\_t0.spec\\
3FGL J0706.5$+$3744 & $<1.04$ & $0.03^{+1.01}_{-0.03}$ & 6.3 & 13.7 & 1.20 & $0.00^{+0.02}_{-0.00}$ & 6.7 & 81.1 & I22491\_60\_TQSO1\_40.sed\\
3FGL J0707.0$+$7741 & $<0.96$ & $0.09^{+0.87}_{-0.09}$ & 7.6 & 19.0 & 1.55 & $0.00^{+0.02}_{-0.00}$ & 11.6 & 91.3 & I22491\_90\_TQSO1\_10.sed\\
3FGL J0753.1$+$5353 & $<1.69$ & $1.63^{+0.06}_{-0.10}$ & 11.2 & 72.7 & 1.90 & $0.00^{+0.05}_{-0.00}$ & 7.2 & 91.4 & M82\_template\_norm.sed\\
3FGL J0754.4$-$1148 & $<2.82$ & $2.73^{+0.09}_{-0.20}$ & 23.0 & 66.0 & 2.00 & $0.01^{+0.05}_{-0.01}$ & 5.8 & 93.8 & Sey2\_template\_norm.sed\\
3FGL J0806.6$+$5933 & $<0.94$ & $0.12^{+0.82}_{-0.12}$ & 10.8 & 23.4 & 1.75 & $0.10^{+0.03}_{-0.03}$ & 15.1 & 93.4 & Sey18\_template\_norm.sed\\
3FGL J0816.7$+$5739 & $<1.22$ & $0.99^{+0.23}_{-0.99}$ & 3.7 & 34.3 & 1.25 & $0.00^{+0.04}_{-0.00}$ & 10.3 & 92.7 & I22491\_70\_TQSO1\_30.sed\\
3FGL J0839.6$+$1803 & $<0.93$ & $0.24^{+0.69}_{-0.24}$ & 22.9 & 26.0 & 1.35 & $0.13^{+0.09}_{-0.04}$ & 16.6 & 82.5 & I22491\_70\_TQSO1\_30.sed\\
3FGL J0843.9$+$5311 & $<1.44$ & $1.08^{+0.36}_{-1.08}$ & 8.6 & 32.6 & 1.35 & $0.00^{+0.01}_{-0.00}$ & 8.2 & 97.9 & I22491\_80\_TQSO1\_20.sed\\
3FGL J0849.3$+$0458 & $<1.14$ & $0.12^{+1.02}_{-0.12}$ & 15.1 & 19.0 & 1.30 & $0.13^{+0.04}_{-0.02}$ & 5.4 & 37.5 & I22491\_40\_TQSO1\_60.sed\\
3FGL J0924.2$+$0534 & $<0.94$ & $0.24^{+0.70}_{-0.24}$ & 4.2 & 25.7 & 0.85 & $0.23^{+0.02}_{-0.08}$ & 2.6 & 64.5 & I22491\_40\_TQSO1\_60.sed\\
3FGL J0942.1$-$0756 & $<1.10$ & $0.06^{+1.04}_{-0.06}$ & 17.6 & 14.6 & 1.75 & $0.00^{+0.02}_{-0.00}$ & 16.8 & 97.0 & I22491\_template\_norm.sed.save\\
3FGL J1027.7$+$6316 & $<1.19$ & $1.06^{+0.13}_{-1.06}$ & 2.8 & 34.3 & 0.85 & $1.43^{+0.02}_{-0.10}$ & 5.6 & 87.1 & pl\_I22491\_30\_TQSO1\_70.sed\\
3FGL J1101.5$+$4106 & $<1.26$ & $1.13^{+0.13}_{-1.13}$ & 0.8 & 36.1 & 0.60 & $0.02^{+0.02}_{-0.02}$ & 3.6 & 14.1 & pl\_I22491\_10\_TQSO1\_90.sed\\
3FGL J1103.9$-$5357 & $...$ & $0.81^{+0.15}_{-0.81}$ & 36.8 & 35.1 & 1.70 & $0.00^{+0.06}_{-0.00}$ & 45.7 & 91.9 & I22491\_template\_norm.sed.save\\
3FHL J1127.8$+$3615 & $<1.65$ & $1.52^{+0.13}_{-0.22}$ & 3.9 & 67.9 & 0.55 & $1.09^{+0.35}_{-0.10}$ & 4.7 & 66.0 & pl\_QSOH\_template\_norm.sed\\
3FGL J1224.6$+$4332 & $...$ & $2.37^{+0.08}_{-0.02}$ & 158.0 & 94.5 & 3.00 & $0.61^{+2.44}_{-0.01}$ & 198.0 & 21.7 & pl\_QSOH\_template\_norm.sed\\
3FGL J1244.8$+$5707 & $<1.06$ & $0.06^{+1.00}_{-0.06}$ & 1.9 & 15.3 & 1.30 & $0.00^{+0.02}_{-0.00}$ & 5.2 & 95.9 & I22491\_70\_TQSO1\_30.sed\\
3FGL J1249.7$+$3705 & $<1.25$ & $1.06^{+0.19}_{-1.06}$ & 0.7 & 33.1 & 0.75 & $0.00^{+0.02}_{-0.00}$ & 2.2 & 27.5 & pl\_I22491\_20\_TQSO1\_80.sed\\
3FGL J1256.7$+$5328 & $<3.39$ & $3.20^{+0.19}_{-0.14}$ & 0.2 & 86.0 & 2.00 & $0.40^{+0.05}_{-0.38}$ & 0.1 & 44.6 & S0\_90\_QSO2\_10.sed\\
3FGL J1311.0$+$0036 & $...$ & $0.93^{+0.01}_{-0.02}$ & 217.0 & 65.9 & 0.35 & $0.29^{+0.02}_{-0.01}$ & 181.0 & 99.6 & pl\_TQSO1\_template\_norm.sed\\
3FGL J1327.9$+$2524 & $<3.66$ & $3.54^{+0.12}_{-0.17}$ & 0.7 & 86.7 & 3.00 & $2.60^{+0.11}_{-0.13}$ & 0.2 & 78.9 & Mrk231\_template\_norm.sed\\
3FGL J1357.5$+$0125 & $<1.09$ & $0.12^{+0.97}_{-0.12}$ & 28.2 & 20.3 & 0.95 & $0.85^{+0.05}_{-0.04}$ & 21.8 & 90.9 & I22491\_80\_TQSO1\_20.sed\\
3FHL J1421.5$-$1654 & $<1.30$ & $0.60^{+0.70}_{-0.60}$ & 13.9 & 24.1 & 1.65 & $0.00^{+0.05}_{-0.00}$ & 15.1 & 74.2 & I22491\_template\_norm.sed.save\\
3FGL J1424.3$+$0434 & $<1.10$ & $0.06^{+1.04}_{-0.06}$ & 2.1 & 14.7 & 1.70 & $0.00^{+0.01}_{-0.00}$ & 8.5 & 95.9 & I22491\_template\_norm.sed.save\\
3FGL J1426.2$+$3402 & $...$ & $0.57^{+0.48}_{-0.55}$ & 370.0 & 29.8 & 3.00 & $0.59^{+0.03}_{-0.03}$ & 557.0 & 99.9 & pl\_QSOH\_template\_norm.sed\\
3FHL J1440.2$-$2343 & $...$ & $2.36^{+0.20}_{-0.00}$ & 63.3 & 71.8 & 1.95 & $0.29^{+0.05}_{-0.01}$ & 3.5 & 77.8 & Mrk231\_template\_norm.sed\\
3FGL J1446.1$-$1628$^f$ & $...$ & $...$ & $...$ & $...$ & $...$ & $...$ & $...$ & $...$ & $...$\\
3FGL J1454.5$+$5124 & $<1.06$ & $0.12^{+0.94}_{-0.12}$ & 1.9 & 20.7 & 1.15 & $0.00^{+0.03}_{-0.00}$ & 4.9 & 89.3 & I22491\_60\_TQSO1\_40.sed\\
3FGL J1509.7$+$5556 & $<1.04$ & $0.03^{+1.01}_{-0.03}$ & 1.9 & 13.7 & 1.20 & $0.00^{+0.03}_{-0.00}$ & 2.1 & 90.6 & I22491\_60\_TQSO1\_40.sed\\
3FGL J1531.0$+$5737 & $...$ & $1.25^{+0.22}_{-1.23}$ & 351.0 & 34.8 & 0.30 & $0.91^{+0.09}_{-0.29}$ & 237.0 & 76.4 & CB1\_0\_LOIII4.sed\\
3FGL J1533.5$+$3416 & $<1.45$ & $1.16^{+0.29}_{-0.32}$ & 9.7 & 40.3 & 0.80 & $1.14^{+0.21}_{-0.05}$ & 10.8 & 76.6 & pl\_QSOH\_template\_norm.sed\\
3FGL J1534.4$+$5323 & $<1.11$ & $0.56^{+0.55}_{-0.56}$ & 1.8 & 27.6 & 0.75 & $0.02^{+0.02}_{-0.02}$ & 1.9 & 24.5 & pl\_I22491\_10\_TQSO1\_90.sed\\
3FGL J1550.3$+$7409 & $<1.65$ & $1.63^{+0.02}_{-1.63}$ & 11.9 & 25.2 & 1.75 & $0.07^{+0.01}_{-0.01}$ & 5.1 & 99.9 & Mrk231\_template\_norm.sed\\
3FGL J1549.0$+$6309 & $<2.94$ & $2.53^{+0.41}_{-0.11}$ & 29.9 & 52.9 & 1.85 & $0.16^{+0.04}_{-0.05}$ & 9.2 & 99.5 & M82\_template\_norm.sed\\
3FGL J1630.7$+$5222 & $<1.22$ & $1.04^{+0.18}_{-1.04}$ & 3.1 & 34.2 & 1.10 & $0.00^{+0.04}_{-0.00}$ & 7.1 & 98.0 & I22491\_60\_TQSO1\_40.sed\\
3FGL J1637.1$+$1314 & $...$ & $0.26^{+0.78}_{-0.26}$ & 238.0 & 23.6 & 0.60 & $1.00^{+0.02}_{-0.04}$ & 133.0 & 65.5 & CB1\_0\_LOIII4.sed\\
3FGL J1651.6$+$7219 & $...$ & $1.04^{+0.13}_{-0.26}$ & 291.0 & 40.7 & 3.00 & $0.59^{+0.03}_{-0.03}$ & 436.0 & 99.9 & pl\_QSOH\_template\_norm.sed\\
3FGL J1829.8$+$1328 & $...$ & $...$ & $...$ & $...$ & $...$ & $...$ & $...$ & $...$ & $...$\\
3FGL J1848.9$+$4247 & $...$ & $0.26^{+0.61}_{-0.26}$ & 134.0 & 27.9 & 0.70 & $0.54^{+0.03}_{-0.03}$ & 77.1 & 99.7 & I22491\_template\_norm.sed.save\\
3FGL J1927.7$+$6118 & $<1.10$ & $0.10^{+1.00}_{-0.10}$ & 1.7 & 18.8 & 1.65 & $0.00^{+0.01}_{-0.00}$ & 12.9 & 92.5 & I22491\_template\_norm.sed.save\\
3FGL J2007.7$-$7728 & $<3.36$ & $3.07^{+0.29}_{-0.55}$ & 0.6 & 64.6 & 0.85 & $3.10^{+0.06}_{-3.10}$ & 0.9 & 55.1 & pl\_QSOH\_template\_norm.sed\\
3FGL J2012.1$-$1643 & $<3.14$ & $2.76^{+0.38}_{-0.11}$ & 3.2 & 80.5 & 0.00 & $2.67^{+0.29}_{-0.02}$ & 2.3 & 84.2 & CB1\_0\_LOIII4.sed\\
3FGL J2107.7$-$4822 & $...$ & $2.98^{+0.27}_{-0.02}$ & 33.9 & 90.0 & 1.95 & $0.16^{+0.05}_{-0.06}$ & 11.7 & 92.0 & Sey2\_template\_norm.sed\\
3FGL J2139.4$-$4235 & $<0.96$ & $0.03^{+0.93}_{-0.03}$ & 2.2 & 14.7 & 1.55 & $0.00^{+0.02}_{-0.00}$ & 8.8 & 97.3 & I22491\_90\_TQSO1\_10.sed\\
3FGL J2149.6$+$1915 & $<1.15$ & $0.82^{+0.33}_{-0.82}$ & 15.8 & 28.2 & 0.80 & $1.12^{+0.08}_{-0.09}$ & 11.3 & 92.3 & pl\_QSOH\_template\_norm.sed\\
3FGL J2159.8$+$1025 & $<2.64$ & $1.90^{+0.74}_{-0.00}$ & 16.4 & 33.3 & 2.00 & $0.05^{+0.06}_{-0.05}$ & 9.2 & 95.6 & M82\_template\_norm.sed\\
3FGL J2236.0$-$3629 & $...$ & $1.93^{+0.12}_{-0.10}$ & 96.5 & 93.4 & 1.70 & $0.03^{+0.03}_{-0.03}$ & 69.5 & 100.0 & Mrk231\_template\_norm.sed\\
3FGL J2236.2$-$5049 & $<2.76$ & $2.62^{+0.14}_{-0.09}$ & 7.5 & 72.9 & 0.85 & $2.57^{+0.08}_{-0.01}$ & 6.3 & 79.5 & pl\_QSOH\_template\_norm.sed\\
3FGL J2240.9$+$4121 & $<1.68$ & $1.53^{+0.15}_{-0.19}$ & 5.2 & 49.1 & 1.20 & $0.00^{+0.02}_{-0.00}$ & 8.1 & 92.2 & I22491\_80\_TQSO1\_20.sed\\
\enddata
    \end{deluxetable*}

\begin{footnotesize}
$^a$ Best-fit or upper limit of the photometric redshift.

$^b$ Photometric redshifts with 1 $\sigma$ confidence level.

$^c$ $\chi^2$ value for ten degrees of freedom.

$^d$  Redshift probability density at $z_{\text {phot }} \pm 0.1\left(1+z_{\text {phot }}\right)$.

$^e$ Spectral slope for the power-law model of the form $F_\lambda \propto \lambda^{-\beta}$.

$^f$ The fittings are not reported because they are not  good ($\chi^2=0$) due to upper limits for six or more filters.
\end{footnotesize}
\\

Following the criteria simulated in \cite{Rajagopal_2020}, sources with $z>1.3$ and $E(B-V) \leq 0.30$ have reliable measured photometric redshifts within $\left|\Delta z /\left(1+z_{\operatorname{sim}}\right)\right|<0.15$ accuracy, where $z_{\operatorname{sim}}$ is the input redshift for the simulation.
Another criterion is the integral of probability density function $P_z > 90\%$, which indicates that the measured redshift is within $0.1(1+z_{\text{phot}})$ of the best-fit value. Reliable upper limits from the power-law model are provided for sources with $P_z \leq 90\% $ and $\chi^2 \leq 30$ for ten degrees of freedom. Specifically, we only report the photometric magnitudes for sources with $E(B-V)>0.3$.

\section{Results}
\begin{figure}[h]
    \centering
    \includegraphics[width=0.7\textwidth]{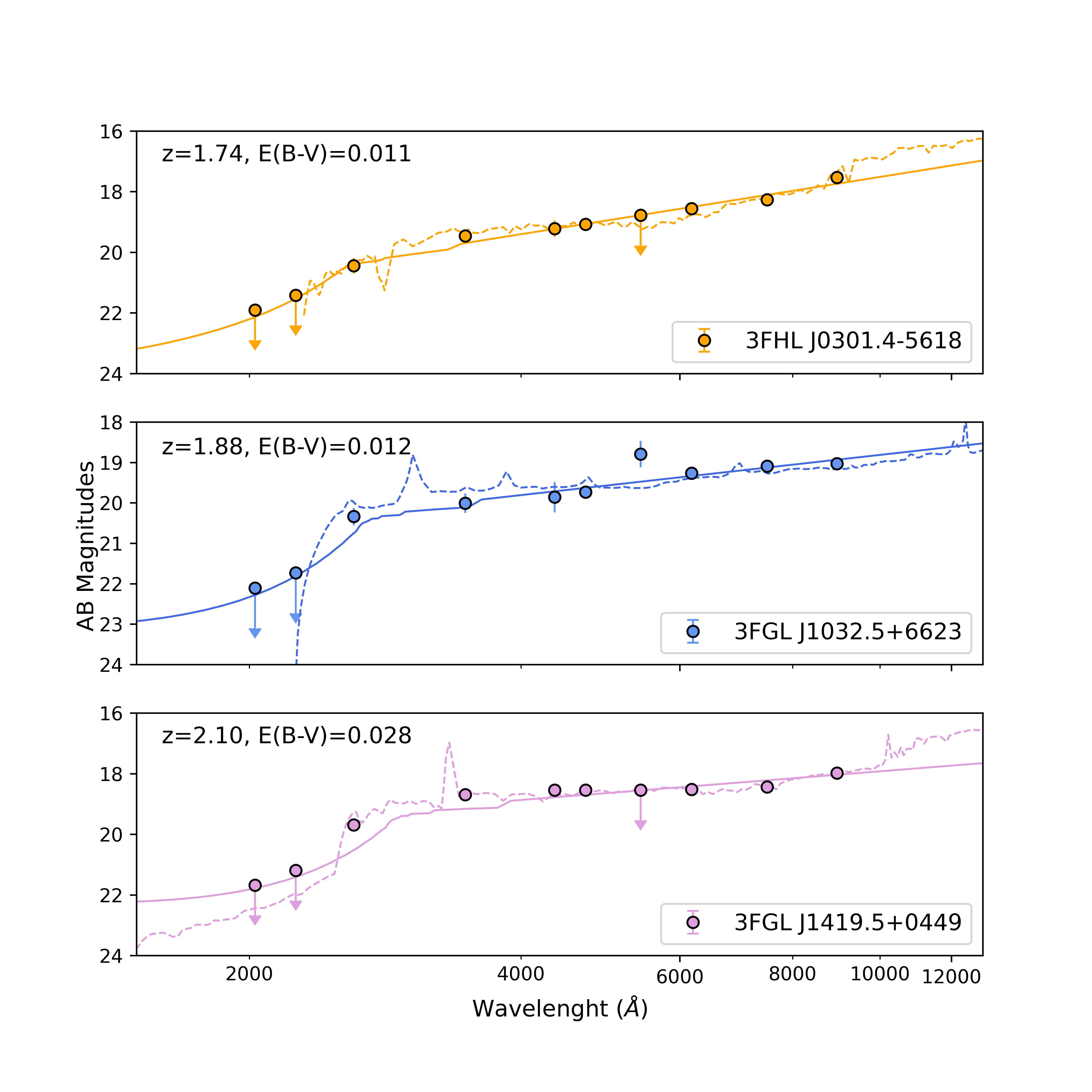}
    \caption{The SEDs of the three high-redshift BL Lacs. The solid circles with black edges represent photometric magnitudes from \emph{Swift}-UVOT, SARA-CT, and SARA-RM. The magnitudes are ordered in the following way: $uw2,\ um2,\ uw1,\ uuu,\ ubb,\ g',\ uvv,\ r',\ i',\ z'$. The solid and dashed lines are power-law and galaxy models, respectively.}
    \label{fig:high-z_sed}
\end{figure}

The best-fit photometric redshifts or upper limits for the BL Lacs, along with the $\chi^2$ and $P_z$ values for the power-law and galaxy template models are reported in Table \ref{tab:result}. Three high-$z$ BL Lacs are found from the analysis of sixty BL Lacs, while forty-one are reported with reliable upper limits. The SEDs in UVOT and SARA filters for the four high-$z$ sources are presented in Figure \ref{fig:high-z_sed}. Under the criteria mentioned above ($P_z \geq 90\%$, \ $E(B-V) \leq 0.3$,\ and $z > 1.3$), the three high-$z$ sources are 3FHL J0301.4$-$5618, 3FGL J1032.5$+$6623, and 3FGL J1419.5$+$0449. The redshifts of the three sources are $1.74^{+0.05}_{-0.08}$, $1.88^{+0.07}_{-0.03}$, and $2.10^{+0.03}_{-0.04}$ for power-law templates, respectively. We also obtained redshifts using galaxy templates at $1.33^{+0.06}_{-0.05}$, $1.49^{+0.39}_{-0.01}$, and $1.73^{+0.01}_{-0.04}$. However, 3FGL J1032.5$+$6623 and 3FGL J1419.5$+$0449 have been identified as BL Lacs in the Fourth Fermi Large Area Telescope Source Catalog (4FGL, \citealt{4fgl}), and therefore the redshifts determined by the galaxy templates are not likely. Although 3FHL J0301.4$-$5618 is unclassified in the 3FGL and 4FGL catalog, it lies in the BL Lac group in the blazar sequence and \emph{Fermi} blazar divide plots (see section 6). Therefore, the redshift determined by the power-law templates is more likely.

\section{Discussions}
\subsection{Cosmic Gamma-ray Horizon}
Very high energy (VHE) photons ($>$100 GeV) do not travel unimpeded. The extragalactic background light (EBL) impedes the travel of VHE photons via photon-photon interactions in the ultraviolet, optical, and infrared bands. Due to these interactions there is a redshift dependent opacity for high-energy photons in the universe \citep{Finke_model, dominguezSpectralAnalysisFermiLAT2015}. This opacity to VHE photons is evaluated by the cosmic gamma-ray horizon (CGRH, \citealt{dominguezDETECTIONCOSMICUpgamma2013}). The CGRH is defined as the energy at which the optical depth is one ($\tau=1$), and it is a function of redshift.

\begin{figure}[h]
\centering
\includegraphics[width=0.7\textwidth]{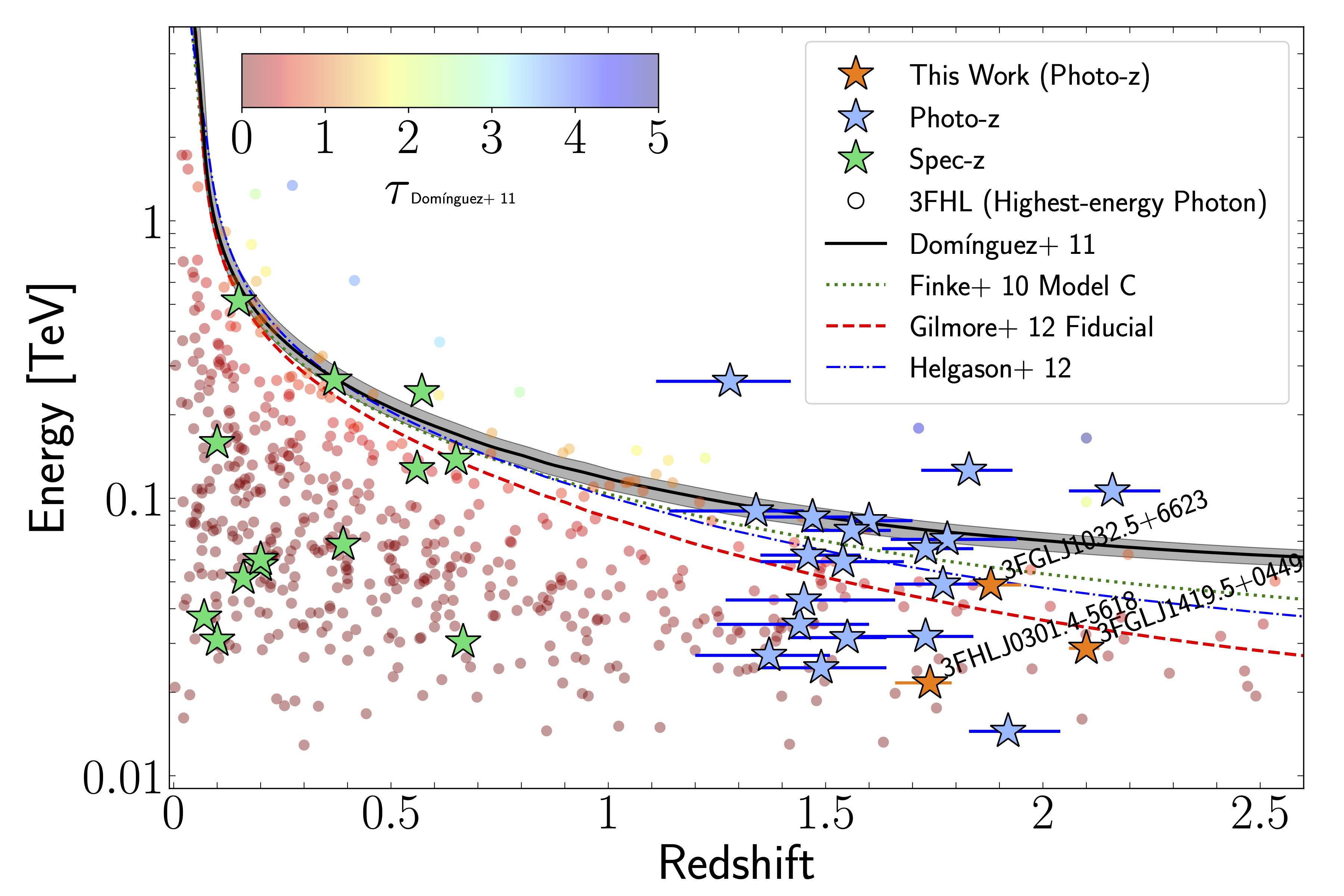}
\caption{The cosmic gamma-ray horizon plot. The colored circles are from the 3FHL catalog. The color gradient shows optical depth. Lines represent the horizon ($\tau$=1) according to different models from \cite{Finke_model}, \cite{Gilmore_CGRH}, \cite{Helgason_CGRH}, and \cite{Dominguez_CGRH}. The stars are the sources with determined redshifts from photometric and spectroscopic methods \citep{marchesi2018identifying,desai2019identifying,rajagopal2021identifying}.}
\label{fig:cgrh}
\end{figure}

Figure \ref{fig:cgrh} shows the highest photon energy versus redshift for different CGRH models. As is made apparent by this figure, there were not many BL Lacs probing the CGRH at high-$z$ before our campaigns. As can been seen, our photometric method is efficient at finding these missing high-$z$ sources. In this work, we find four high-$z$ sources: 3FGL J1419.5$+$0449 ($\tau=0.16_{-0.03}^{+0.05}$,\  $E_{\rm max}=29\ \mathrm{GeV}$), 3FGL J1032.5$+$6623 ($\tau=0.44_{-0.08}^{+0.12}$,\ $E_{\rm max}=49\ \mathrm{GeV}$), 3FHL J0301.4$-$5618 ($\tau=0.05_{-0.01}^{+0.02}$, \ $E_{\rm max}=22\ \mathrm{GeV}$). All of our sources have $\tau<1$, which means that they are consistent with the CGRH models developed by \cite{Dominguez_CGRH} and \cite{Finke_model}.

\begin{figure}[h]
    \centering
    \includegraphics[width=0.5\textwidth]{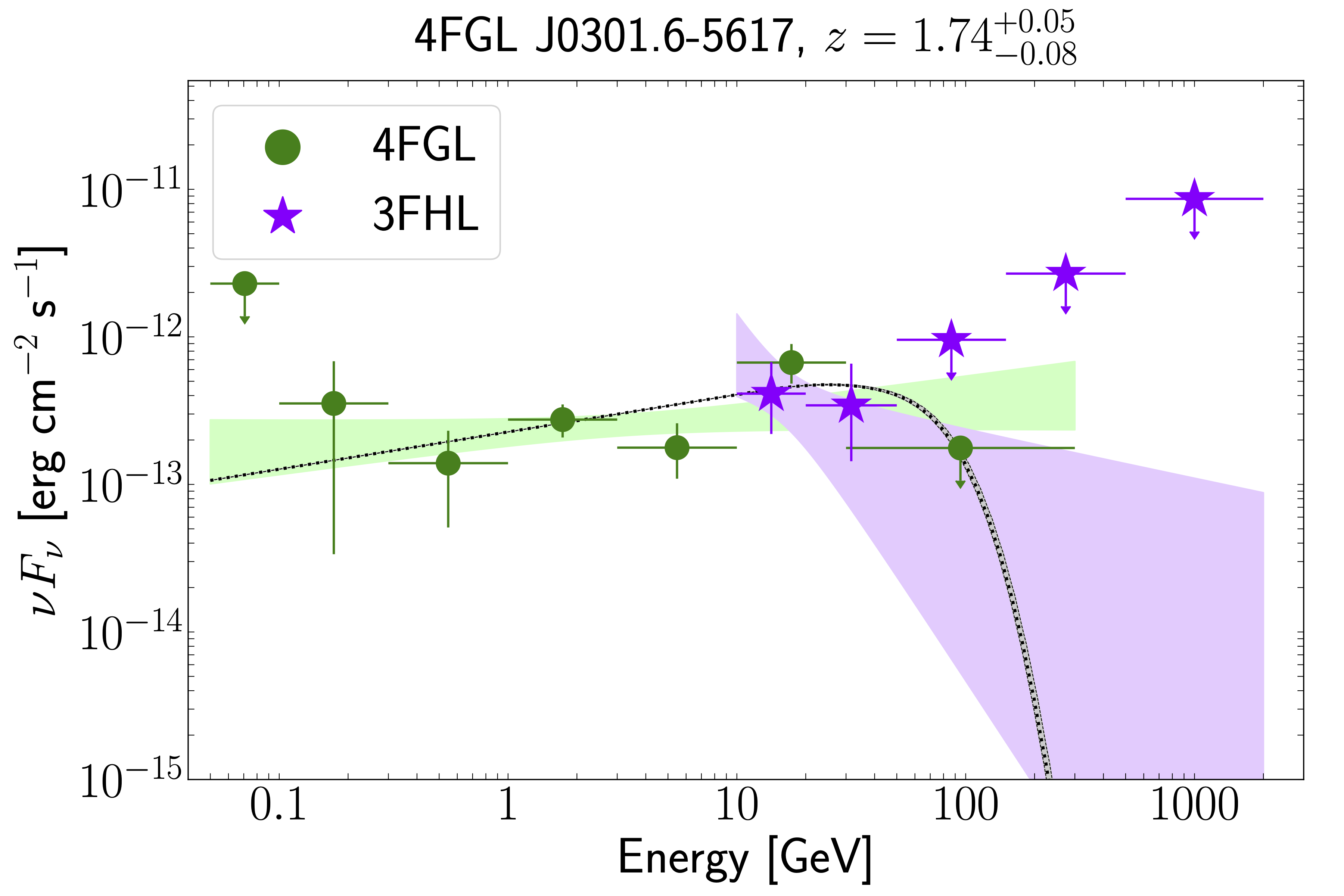}\hfill
    \includegraphics[width=0.5\textwidth]{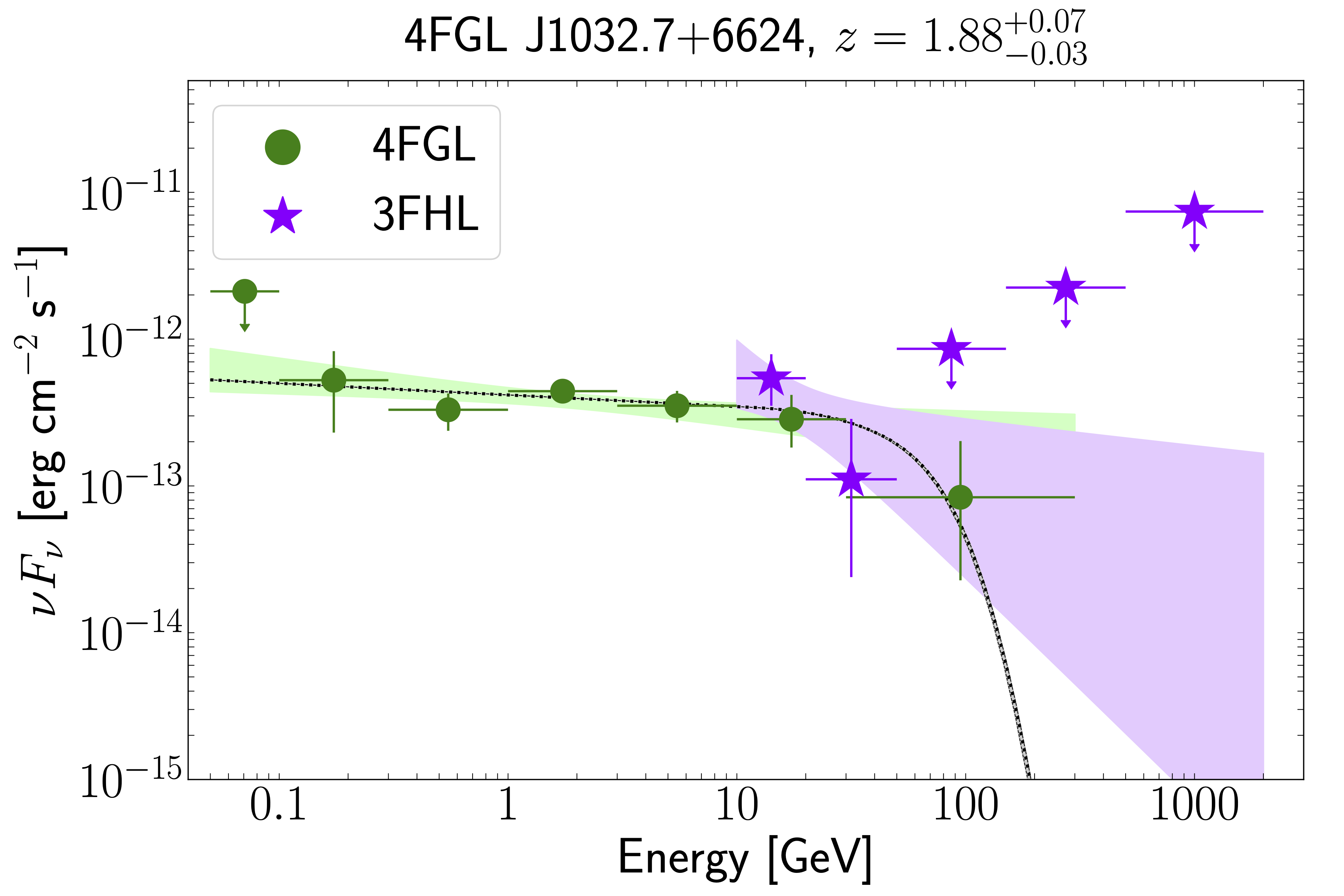}\hfill
    \includegraphics[width=0.5\textwidth]{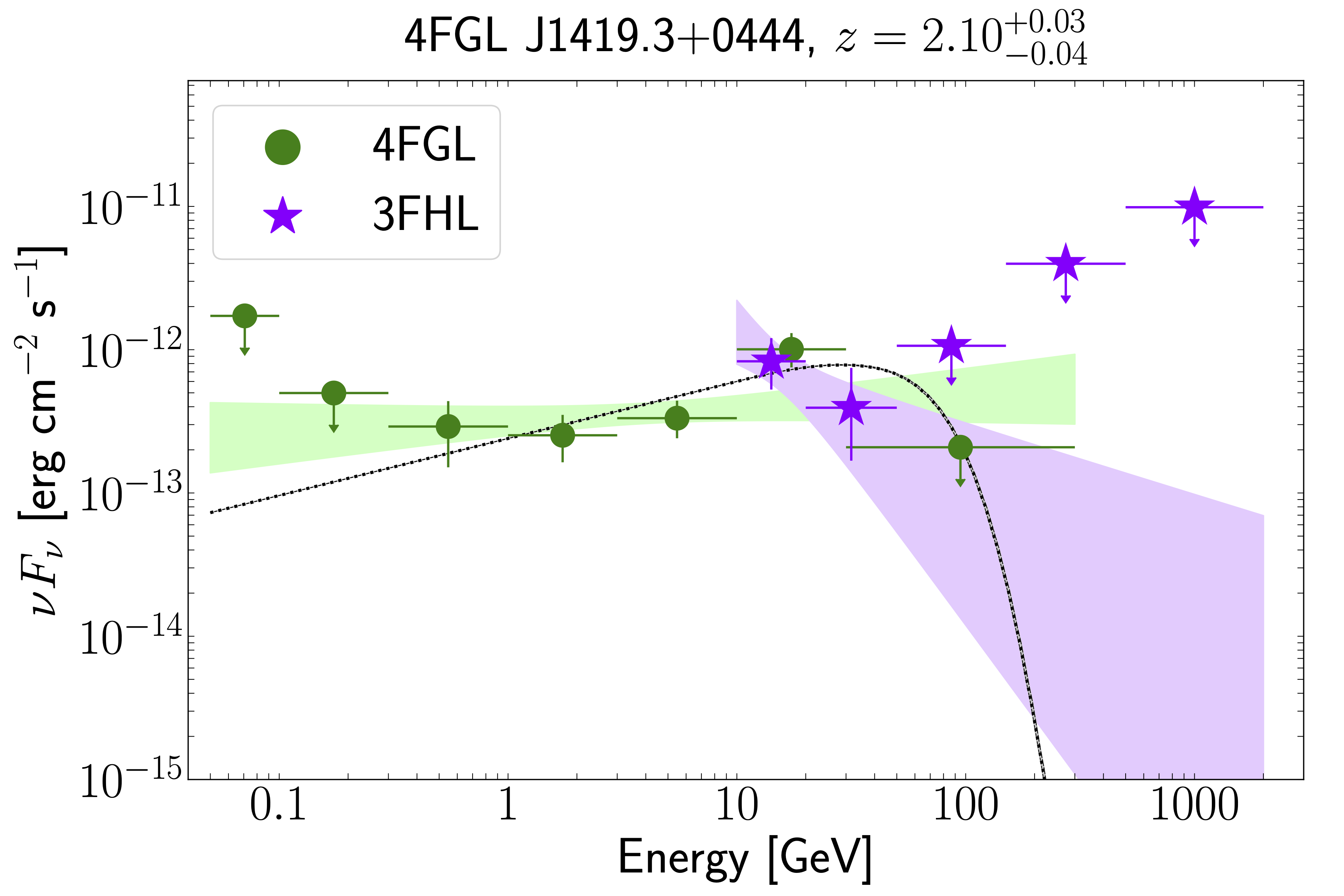}\hfill
    \caption{The SED data points are from the 4FGL (green dots) and 3FHL (purple stars) catalogs. The green and purple areas are the power-law fits with 1 $\sigma$ uncertainty provided in the 4FGL and 3FHL catalogs. The green and purple uncertainty bands come from the \emph{Fermi}-LAT data reduction and include statistical uncertainties The black curves are the best joint power-law fits of 4FGL and 3FHL data attenuated by a factor $e^{-\tau(E,z)}$ due to EBL absorption. The EBL model is taken from \cite{Dominguez_CGRH}. The uncertainty region (the grey shaded area) shown for this joint fit is exclusively coming from the redshift uncertainty, according to \cite{Dominguez_CGRH}, and the gray area is not easily visible due to the small effects from the redshift uncertainty. The joint fit shows where the energy cutoff is. Note that the energy cutoff is solely due to the EBL attenuation.}
    \label{fig:sed}
    
\end{figure}

The gamma-ray SEDs of the four high-$z$ BL Lacs are shown in Figure \ref{fig:sed}. We fit the Fermi-LAT Fourth Source Catalog (4FGL, \citealt{4fgl,4fgl_dr2}) and 3FHL data with a power law  attenuated by EBL absorption. The EBL absorption \citep{Dominguez_CGRH} is applied with a factor $e^{-\tau(E,z)}$ in the fits. The redshift uncertainty determined by the $\chi^2$ fitting is considered in the fits, which is demonstrated by the gray shaded area in the plots. The fits show that redshift uncertainty has a negligible effect on the source flux. Moreover, the effect of redshift uncertainty decreases with higher redshift. In addition, the attenuation caused by the EBL can be seen in the joint fits of the 3FHL and 4FGL data.

\subsection{Blazar Sequence}
\cite{fossati1998unifying} introduced the blazar sequence to find a unified model for all types of blazars. The blazar sequence is based on three relations among the physical properties of blazars \citep{prandini2022blazar}. The first is an anti-correlation between the rest frame synchrotron peak frequency ($\nu^{\mathrm{sy}}_{\mathrm{pk,\ rest}}$) and the luminosity ($L_{pk}^{sy}$) at this peak frequency. FSRQs occupy the region with lower $\nu^{\mathrm{sy}}_{\mathrm{pk,\ rest}}$ and higher $L_{pk}^{sy}$. On the contrary, BL Lacs show higher $\nu^{\mathrm{sy}}_{\mathrm{pk,\ rest}}$ and lower $L_{pk}^{sy}$. However, this phenomenon could also be ascribed to selection effects \citep{giommi2012simplified}. BL Lacs with high $\nu^{\mathrm{sy}}_{\mathrm{pk,\ rest}}$ and high $L_{pk}^{sy}$ could exist among those that do not have a redshift measurement available.
Thus, measuring the redshift of BL Lacs is important to constrain the blazar sequence. The second relation is the anti-correlation between $\nu^{\mathrm{sy}}_{\mathrm{pk,\ rest}}$ and Compton dominance. Compton dominance is defined as the ratio of the inverse-Compton peak luminosity ($L_{pk}^{IC}$) to $L_{pk}^{sy}$. BL Lacs and FSRQs populate different regions of this plane: FSRQs have higher Compton dominance compared to BL Lacs. The last relation is also an anti-correlation, but it is between the spectral index ($\Gamma_{\gamma}$) and $\nu^{\mathrm{sy}}_{\mathrm{pk,\ rest}}$. They follow similar behavior as the second one mentioned above: FSRQs have larger gamma-ray index and populate lower synchrotron frequencies, while BL Lacs have smaller gamma-ray index and populate higher synchrotron frequencies.  

%
%
%

\begin{figure}[!htb]
    \centering
    \subfigure[Synchrotron peak luminosity vs. rest-frame synchrotron peak frequency.]
    { \includegraphics[width=0.48\textwidth]{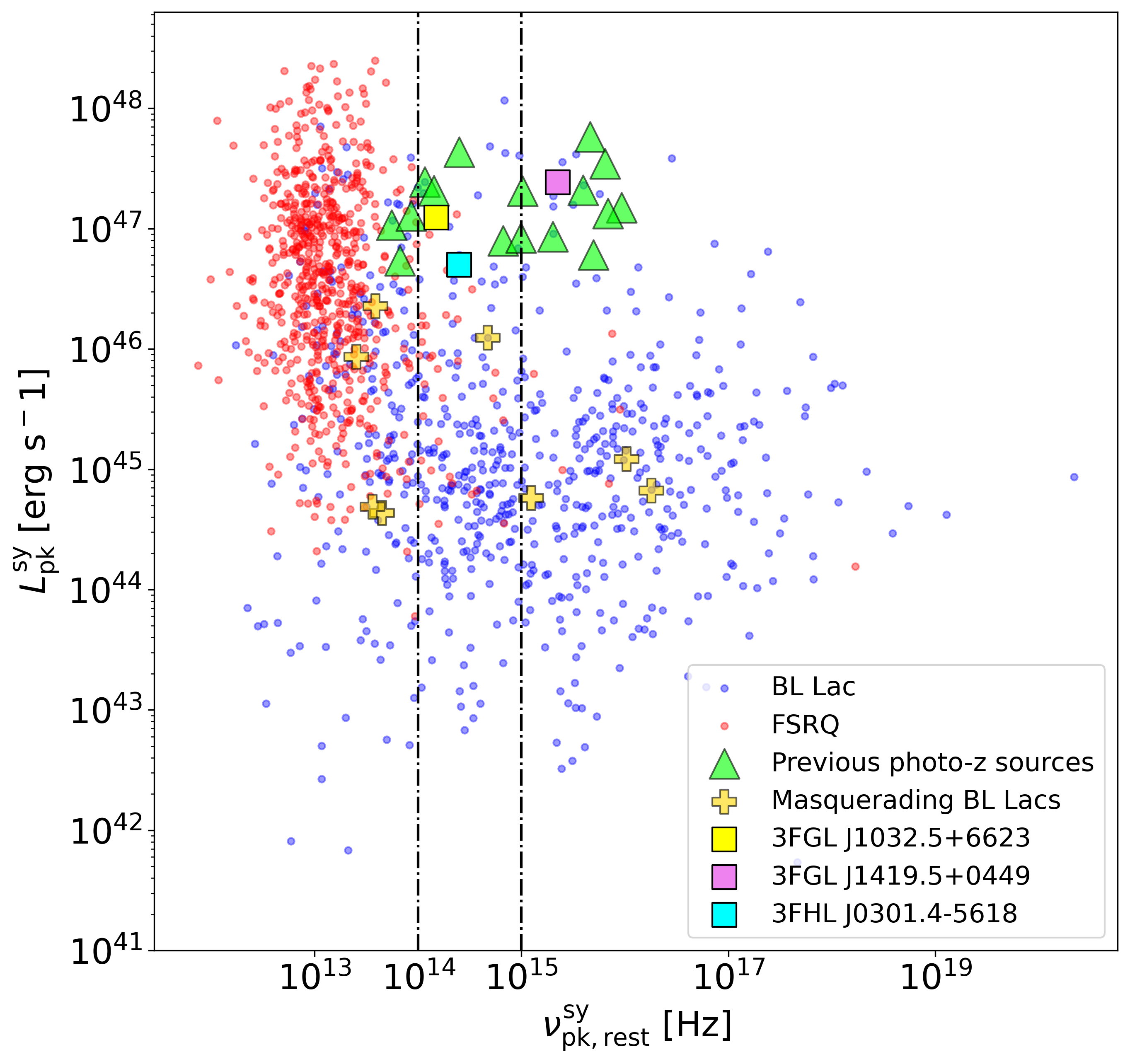}
    \label{fig:sequence_syn_lumi} }\hfill
    \subfigure[Compton dominance vs. synchrotron peak frequency at rest frame.]
    { \includegraphics[width=0.48\textwidth]{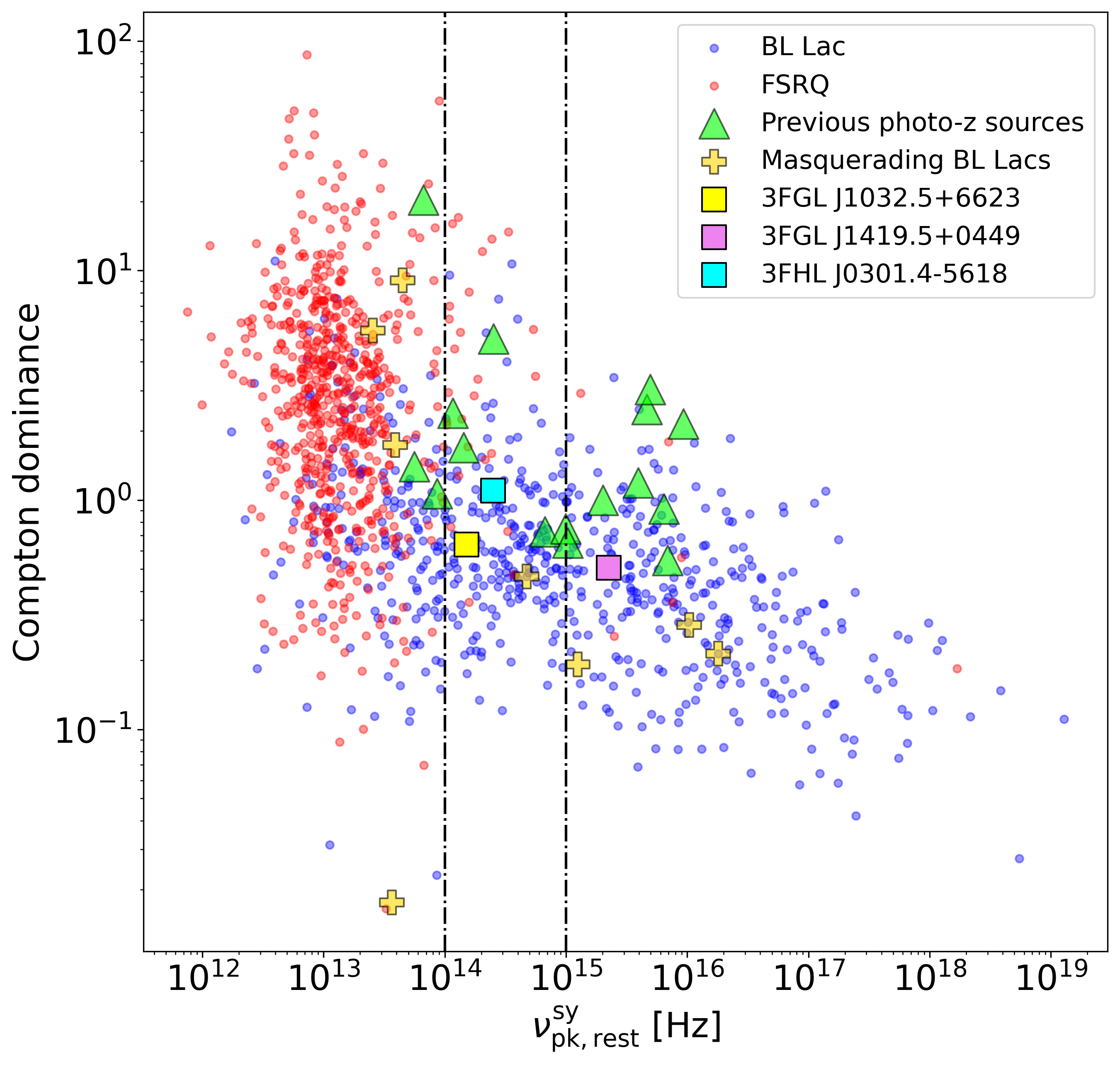}
    \label{fig:sequence_cd} }\\
    \subfigure[Gamma-ray index vs. synchrotron peak frequency at rest frame.]
    { \includegraphics[width=0.48\textwidth]{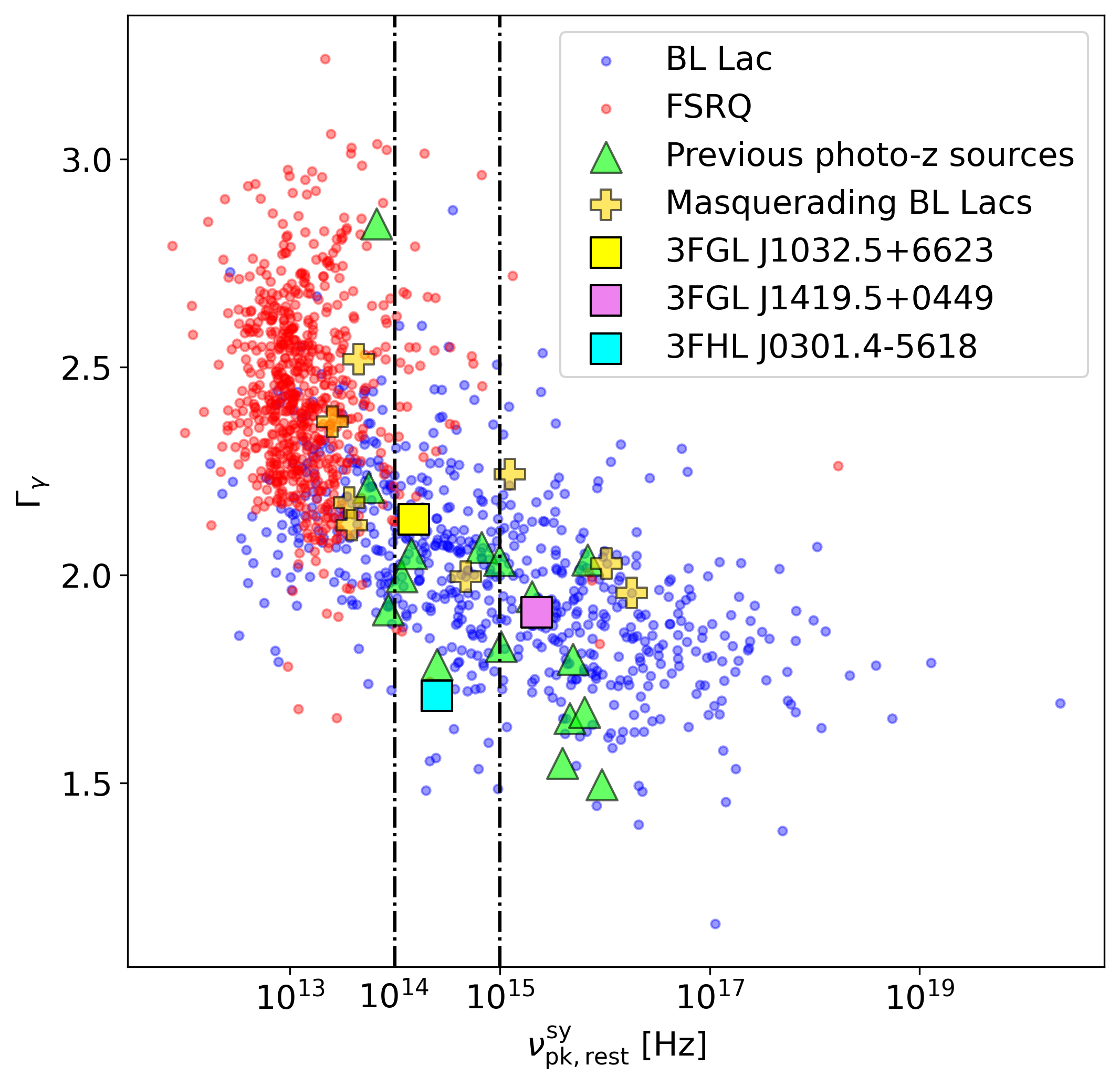}
    \label{fig:sequence_index} }
    \subfigure[Gamma-ray index vs. gamma-ray luminosity (0.1-100 GeV).]
    { \includegraphics[width=0.48\textwidth]{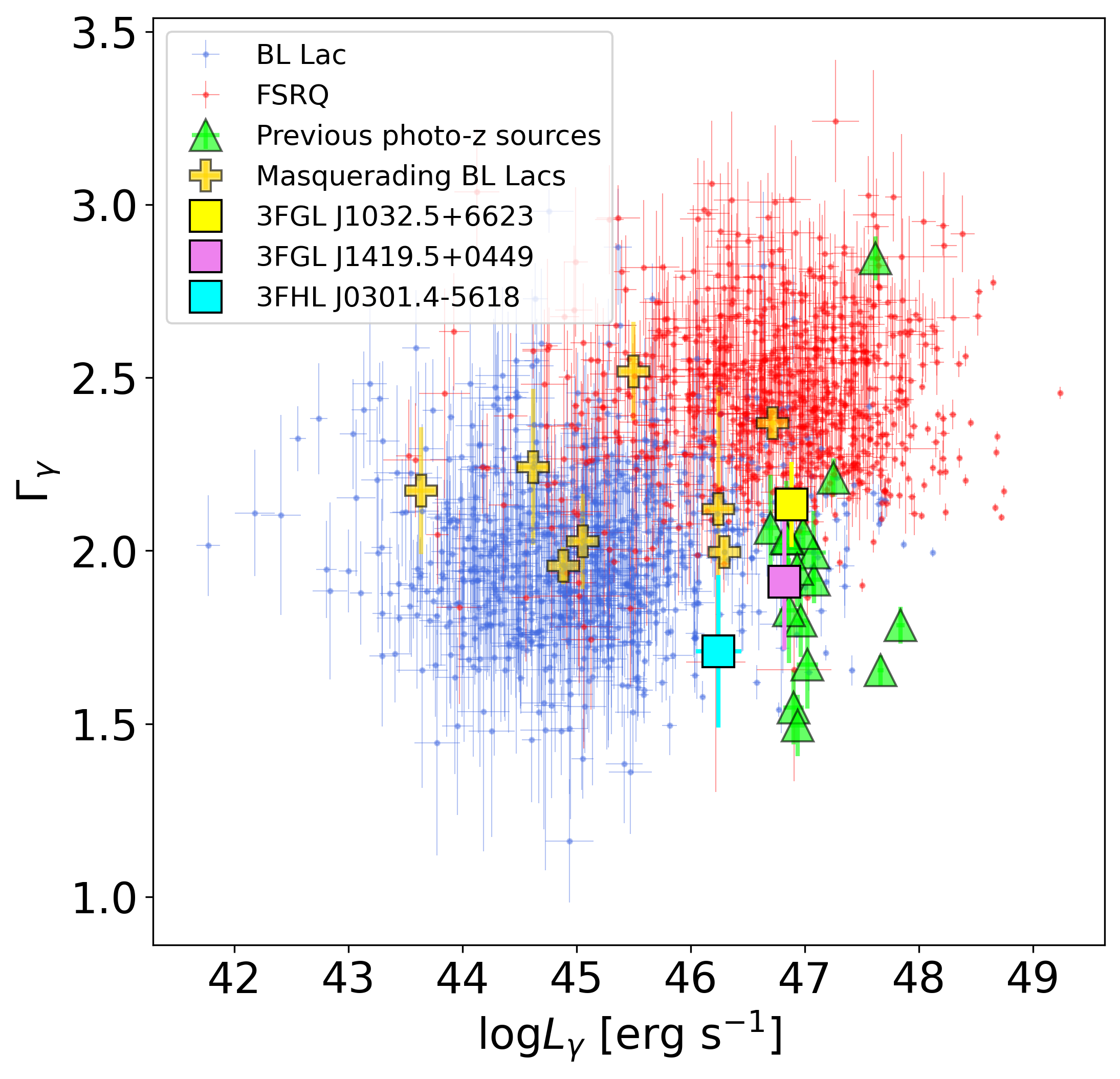}
    \label{fig:blazar_divide} }
\caption{The colored dots are from the 4LAC catalog \citep{Ajello_2020}. The triangles represent the sources from previous photo-z papers \citep{Rau2012,kaur2017,kaur2018,Rajagopal_2020}, while the squares are found in this work. The vertical dotted lines are the divisions for LSP, ISP, and HSP blazars. The orange crosses are the masquerading BL Lacs from \cite{mbl_lacs} .Note that the indices are not corrected for EBL absorption.} 
\label{fig:sequence_divide}
\end{figure}




In order to explore these relations, we use the SSDC Sky Explorer\footnote{https://tools.ssdc.asi.it/} to calculate the synchrotron peak frequency and flux of high-$z$ sources if they are not provided in the 4LAC catalog. We also use the equation from \cite{lea_blazar_compton} to calculate the inverse Compton peak flux for sources with power-law gamma-ray spectra:

\begin{equation} \label{eq2}
\frac{d N}{d E}=K\left[\left(\frac{E}{E_{b}}\right)^{\delta_{1}}+\left(\frac{E}{E_{b}}\right)^{\delta_{2}}\right]^{-1}
\end{equation}
The break energy $E_{b}$ is estimated from the power-law photon index $\Gamma$  using the $E_{b}-\Gamma$ relationship reported in \cite{ajello_ep_index}. The normalization constant $K$ is set to reproduce the  1-100\ GeV flux reported in 
 the 4FGL catalog. Figure \ref{fig:sequence_syn_lumi} shows the existence of high-$z$ luminous BL Lacs that are not detected easily via traditional optical spectroscopy but are discovered by our photometric redshift.
 On the other hand,  in both Figures \ref{fig:sequence_cd} and  \ref{fig:sequence_index},
 our objects are more in line with the standard regions occupied by BL Lacs.

\subsection{The Fermi Blazar Divide}
After three months of Fermi-LAT observations \citep{abdo2009bright}, \cite{ghisellini2009fermi} reported a clear divide between BL Lacs and FSRQs on the gamma-ray index ($\Gamma_{\gamma}$) and gamma-ray luminosity ($L_{\gamma}$) plane. BL Lacs are less luminous ($L_{\gamma}<10^{47}\ \mathrm{erg\ s^{-1}}$) and have harder spectra ($\Gamma_{\gamma}<$2.2) than FSRQs. Different accretion rates might account for this phenomenon \citep{ghisellini2009fermi}. Figure \ref{fig:blazar_divide} reproduces the \emph{Fermi} blazar divide plot using the 4LAC catalog \citep{Ajello_2020} and the high-$z$ sources found by our photometric method. We calculate the gamma-ray luminosity from 0.1 GeV to 100 GeV using equation 1 in \cite{ghisellini2009fermi}. All the high-$z$ sources this paper find have a harder spectrum ($\Gamma_{\gamma} < 2.2$). Although there is an overlap between BL Lacs and FSRQs, the separation of the two populations is still distinct. However, our BL Lac objects have properties in between FSRQs and BL Lacs (they have high luminosities and hard spectra).

 \cite{giommi2013simplified} proposed that these BL Lacs are ``blue quasars" or ``masquerading BL Lacs" because the overwhelming synchrotron emission outshines their broad emission lines.
%
%
According to \cite{2012MNRAS}   the luminosity of the broad line region is connected to the gamma-ray luminosity by $L_{BLR} \sim 4L_{\gamma}^{0.93}$. On average we assume that 10\% of the disk luminosity ($L_{\mathrm{disk}}$) is processed by the broad line region: $L_{\mathrm{BLR}}=0.1L_{\mathrm{disk}}$. Then the ratio of $L_{\mathrm{disk}}$ to $L_{\mathrm{Edd}}$ can be found. For the four high-$z$ sources in this work, the ratios are 0.027, 0.023, and 0.007 for 3FGL J1032.5$+$6623, 3FGL J1419.5$+$0449, and 3FHL J0301.4$-$5618 respectively. Two of the three sources have the ratio between 0.02 to 0.05, suggesting efficient accretion as typical for FSRQs. The overall properties of the BL Lacs with high photometric redshifts ($z>1.3$) allow us to characterize them as potential masquerading BL Lacs, thus increasing this interesting population.


\section{Summary and Conclusions}
In this work we use 10-filter photometry with {\it Swift}-UVOT and SARA to constrain the redshift of 60 BL Lac objects without a spectroscopic measurement. Our method, based on the method used in \cite{Rau2012},  can produce a photometric redshift measurement for BL Lacs at $z \geq 1.3$ or an upper limit for BL Lacs at lower redshift. Out of the 60 BL Lacs, we discovered 3 new high-$z$ ($z \geq$1.3) objects and provided upper limits for 41. For the remaining 15 objects, our fits to the photometric data failed to provide an acceptable solution. This work, together with the previous three in this photometric campaign, have discovered a total of 19 BL Lacs with a $z>1.3$ photometric redshift. 
 Increasing the number of high-$z$  BL Lacs is crucial since they are rare but essential in studying the blazar population and constraining the EBL. Our work shows that these objects have characteristics (high luminosity and high synchrotron peak frequencies) in between those of FSRQs and BL Lacs and that they could belong to the elusive class of `masquerading BL Lacs'. Moreover, we find that the photons detected from these objects probe the high-$z$ portion of the CGRH. We also find that the redshift uncertainty has negligible impact on the modeling of the EBL, allowing the use of these newly discovered high-$z$ BL Lacs in future measurements of the EBL.

\acknowledgements

Y.S. and M.A. acknowledge funding under NASA contracts 80NSSC21K1888 and 80NSSC21K1400. The authors acknowledge the \emph{Swift} team for scheduling all the \emph{Swift}-UVOT observations. They also acknowledge the Southeastern Association for Research in Astronomy for providing SARA-CT and SARA-RM observations. A.D. is thankful for the support of the Ramón y Cajal program from the Spanish MINECO.

\software{
    Astropy (\citealp{astropy}),\ 
    HEASoft (v6.28; \citealp{heasoft}),\ 
    numpy (\citealp{numpy}),\ 
    SAO Image DS9 (\citealp{ds9}),\ 
    Matplotlib (\citealp{matplotlib}),\ 
    Pandas (\citealp{pandas}),\ 
    Scipy (\citealp{sciPy})
    }

\bibliographystyle{apj}
\bibliography{SARA_Photo_z}

\end{document}